\RequirePackage[2020-02-02]{latexrelease}
\documentclass{article} 

\usepackage{preprint}

\usepackage{amsmath, amssymb}
\usepackage{siunitx}
\usepackage[super]{nth}

\usepackage[utf8]{inputenc}	
\usepackage[T1]{fontenc}	
\usepackage{xcolor}		
\usepackage[colorlinks = true,
            linkcolor = purple,
            urlcolor  = blue,
            citecolor = cyan,
            anchorcolor = black]{hyperref}	
\usepackage{booktabs} 		
\usepackage{nicefrac}		
\usepackage{float}			
\usepackage{multicol}		

\usepackage{newfloat}
\DeclareFloatingEnvironment[name={Supplementary Figure}]{suppfigure}
\usepackage{sidecap}
\sidecaptionvpos{figure}{c}

\usepackage{titlesec}
\titlespacing\section{0pt}{12pt plus 3pt minus 3pt}{1pt plus 1pt minus 1pt}
\titlespacing\subsection{0pt}{10pt plus 3pt minus 3pt}{1pt plus 1pt minus 1pt}
\titlespacing\subsubsection{0pt}{8pt plus 3pt minus 3pt}{1pt plus 1pt minus 1pt}

\usepackage{multirow}
\usepackage{graphicx}

\usepackage{interval}
\intervalconfig{
soft open fences,
}

\newcommand\SecNum[1]{%
  \hyperref[#1]{\getrefnumber{#1}}%
}

\usepackage{caption}
\usepackage{subcaption}

\usepackage{placeins}

\usepackage{tikz,xcolor}

\definecolor{lime}{HTML}{A6CE39}
\DeclareRobustCommand{\orcidicon}{
	\begin{tikzpicture}
	\draw[lime, fill=lime] (0,0) 
	circle [radius=0.16] 
	node[white] {{\fontfamily{qag}\selectfont \tiny ID}};
	\draw[white, fill=white] (-0.0625,0.095) 
	circle [radius=0.007];
	\end{tikzpicture}
	\hspace{-2mm}
}
\foreach \x in {A, ..., Z}{\expandafter\xdef\csname orcid\x\endcsname{\noexpand\href{https://orcid.org/\csname orcidauthor\x\endcsname}
			{\noexpand\orcidicon}}
}

\title{Yield and Buckling Stress Limits in Topology Optimization of Multiscale Structures}

\usepackage{xwatermark}
\newwatermark[firstpage,color=gray!90,angle=0,scale=0.28, xpos=0in,ypos=-5in]{*correspondence: \texttt{chrify@dtu.dk}}

\usepackage{authblk}

\author[1\thanks{\tt{chrify@dtu.dk}}]{Christoffer Fyllgraf Christensen\orcidA{}}
\author[1]{Fengwen Wang\orcidB{}}
\author[1]{Ole Sigmund\orcidC{}}

\affil[1]{Department of Civil and Mechanical Engineering, Technical University of Denmark, Koppels Alle 404, 2800 Kgs. Lyngby, Denmark.}

\usepackage{array}
\usepackage{makecell}
\usepackage{gensymb}
\usepackage{appendix}
\usepackage{xurl}

\usepackage{placeins}

\usepackage{lmodern}

\usepackage{bm}

\usepackage{booktabs}

\usepackage{pgfplots}
\usetikzlibrary{plotmarks}
\usetikzlibrary{arrows.meta}
\usetikzlibrary{patterns}
\usetikzlibrary{patterns.meta} 
\usetikzlibrary{positioning}
\usetikzlibrary{decorations.markings}

\usepgfplotslibrary{patchplots}

\usetikzlibrary{external}

\pgfplotsset{plot coordinates/math parser=false}
\newlength\figureheight
\newlength\figurewidth 

\pgfplotsset{yaxis stuff style/.style={y axis line style = {#1},
y tick label style= {#1},
y tick style= {#1},
ylabel style = {#1},
}}

\definecolor{mycolor1}{rgb}{0.384313725,0.6,0.745098039}%
\definecolor{mycolor2}{rgb}{0.8314,0.2784,0.2471}%
\definecolor{mycolor3}{rgb}{0.980392157,0.658823529,0.509803922}
\definecolor{mycolor4}{rgb}{0.4, 0.4, 0.6}
\definecolor{mycolor5}{rgb}{0,0.62,0.451}%
\definecolor{mycolor6}{rgb}{0.8,0.475,0.655}%
\definecolor{mycolor7}{rgb}{0.941,0.894,0.258}%

\usepackage{multirow}

\usepackage[normalem]{ulem}

\usepackage{lineno} 
\let\oldequation\equation
\let\oldendequation\endequation
\let\oldalign\align
\let\oldendalign\endalign

\usepackage{cleveref}

\renewenvironment{equation}
  {\linenomathNonumbers\oldequation}
  {\oldendequation\endlinenomath}
\renewenvironment{align}
  {\linenomathNonumbers\oldalign}
  {\oldendalign\endlinenomath}

\newcolumntype{L}{>{\raggedright\arraybackslash}X}

\newenvironment{Figure}
  {\par\medskip\noindent\minipage{\linewidth}}
  {\endminipage\par\medskip}

\newenvironment{Table}
  {\par\medskip\noindent\minipage{\linewidth}}
  {\endminipage\par\medskip}

\begin{document}


\maketitle

\begin{abstract}
    This study presents an extension of multiscale topology optimization by integrating both yield stress and local/global buckling considerations into the design process. Building upon established multiscale methodologies, we develop a new framework incorporating yield stress limits either as constraints or objectives alongside previously established local and global buckling constraints. This approach significantly refines the optimization process, ensuring that the resulting designs meet mechanical performance criteria and adhere to critical material yield constraints. First, we establish local density-dependent von Mises yield surfaces based on local yield estimates from homogenization-based analysis to predict the local yield limits of the homogenized materials. Then, these local Yield-based Load Factors (YLFs) are combined with local and global buckling criteria to obtain topology optimized designs that consider yield and buckling failure on all levels. This integration is crucial for the practical application of optimized structures in real-world scenarios, where material yield and stability behavior critically influence structural integrity and durability. Numerical examples demonstrate how optimized designs depend on the stiffness to yield ratio of the considered building material. Despite the foundational assumption of separation of scales, the de-homogenized structures, even at relatively coarse length scales, exhibit a high degree of agreement with the corresponding homogenized predictions.
\end{abstract}

\vspace{0.35cm}

\begin{multicols}{2}

    \section{Introduction}
    \label{sec:intro}
    \noindent Topology optimization has become an increasingly essential tool in mechanical engineering, since the pioneering work on homogenization-based optimization by Bendsøe and Kikuchi \cite{BENDSOE1988197}. It enables the creation of optimized structural layouts within a defined physical domain based on specific objectives and constraints. Its primary advantage lies in its ability to generate innovative and efficient designs with minimal prior knowledge, making it highly valuable for industrial applications.
    
    Since the begining, topology optimization has evolved with various approaches \cite{Sigmund2013}, including the widely used singlescale density approach, popular due to its simplicity and effectiveness. Achieving high resolution in practical designs requires a large number of elements, leading to significant computational costs. For practical design problems, achieving high resolution in topology optimization necessitates at least mega- or even giga-scale element counts, as demonstrated by Aage et al. \cite{Aage2017} and Baandrup et al. \cite{Baandrup2020}. This offers new insights and perspectives on structural design such as revealing advantage of multiscale structure. 
    
    Structures with multiscale material can achieve superior mechanical properties such as increased stiffness per weight, strength, and toughness by optimizing material distribution at multiple scales, allowing for lightweight yet high-performance designs \cite{Wu2021}. Tailoring the microstructure enables control over properties like thermal expansion, conductivity, and acoustic damping, which are critical in various engineering fields. Additionally, multiscale structures improve structural integrity by enhancing resistance to failure mechanisms such as buckling, yielding, and fatigue, leading to more reliable and durable designs \cite{Clausen2016,Wu2018,Christensen2025}. They also allow for functional integration, combining load-bearing capacity with other functions like thermal insulation \cite{Yan2015,Jin2023,Sun2024}, thus reducing the need for additional materials. Another advantage of multiscale approaches is that computational cost may be significantly reduced by assuming separation of scales and use of advanced homogenization techniques.
    
    Reducing computational complexity while maintaining high-resolution structural designs remains an ongoing research topic. A promising approach is the de-homogenization procedure proposed by Pantz and Trabelsi \cite{pantz2008a} and subsequently improved by Groen and Sigmund \cite{Groen2018} with a noticeable contribution on how to handle singularities during de-homogenization by Geoffroy-Donders \cite{Geoffroy-Donders2018}. Rescently the method was extended to multiple load cases by Jensen et al. \cite{Jensen2022}. This multiscale design method begins with a homogenization-based topology optimization \cite{BENDSOE1988197}, followed by mapping these microstructures to a larger domain with limited performance loss through de-homogenization.
    
    In standard topology optimization, considerations of buckling and yield stress constraints are crucial but still represent a challenge c.f. da Silva et al. \cite{DaSilva2021} and Ferrari and Sigmund \cite{ferrari2019a}. However, in multiscale topology optimization, these considerations become even more complex due to the multiple scales involved. Despite this complexity, multiscale buckling design can yield better performing structures, as demonstrated by Clausen et al. \cite{Clausen2016}. Recent research by Christensen et al. \cite{Christensen2023} and Hübner et al. \cite{Hubner2023,Hubner2023a} has shown how buckling constraints can be incorporated into the multiscale design process. However, the additional integration of yield stress constraints into multiscale topology optimization has yet to be demonstrated. Multiscale stress-constraints alone where considered in the work by Duysinx and Bendsøe \cite{duysinx1998a}. Early work by Yuge and Kikuchi \cite{yuge1995a} focused on frame structures considering plasticity on both scales using a square unit cell with a variable-sized square hole. Recent work by Wei et al. \cite{Wei2023} looked at concurrent optimization of macro- and microscale structures with yield constraints. This enabled designs optimized againts yield failiure on both scales, but was limited to a fixed number of microstructures with discrete jumps in volume fraction.
    
    Until recently, yield and buckling constraints have been dealt with separately in singlescale topology optimization. The work by Russ and Waisman \cite{Russ2021} combined buckling and yield constraints in singlescale optimization. Work by Wang et al. \cite{Wang2021} focused on using filleted lattice structures considering a simplified local buckling formulation and yield stress constraints for compliance optimization.

    \begin{figure*}[tb]
        \centering
        \graphicspath{{figures/multiscaleIllustration/}}
        \makebox[\textwidth][c]{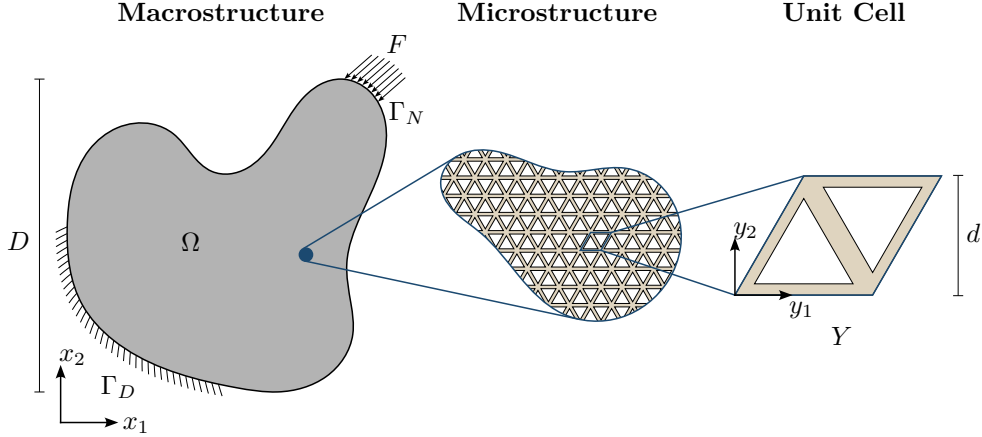}%
        \caption{Illustration of the separation of scales of an isotropic multiscale structure for topology optimization: The macrostructure in the global frame $\bm{x}$  and the microstructure in the local frame $\bm{y}$ which is described by a repetition of the unit cell.}
        \label{fig:multiscaleIllustration}
    \end{figure*}
    The present study aims to solve the standing challenge of combined multiscale yield failure and multiscale buckling stability in one unified topology optimization framework. This work builds upon and enhances the methodologies presented in the PhD thesis by Christensen \cite{Christensen2024}. The approach used here considers a macro- and a microscale and assumes a perfectly periodic microstructure with a clear separation of scales. This is illustrated in \cref{fig:multiscaleIllustration}. Where the material modeled on the macroscale is heterogeneous in the global frame $\bm{x}$, with a characteristic length $D$ equivalent to the size of the design domain. At the microscale level, the material has a characteristic length $d$ corresponding to the size of the unit cell, and it is assumed homogeneous in the local frame $\bm{y}$. The separation of scale can be expressed as $\epsilon = d/D \in \interval[open left]{0}{1}$. For an extensive review of multiscale topology optimization, the reader is referred to Wu et al. \cite{Wu2021}. 
    
    In contrast to the work by Christensen et al. \cite{Christensen2023}, which utilized a triangular microstructure with perfectly sharp corners (as shown in \cref{fig:multiscaleIllustration}), this study develops an improved microstructure featuring smoothed boundaries. This approach mitigates designs purely driven by microscale stress concentrations. The new microstructure resembles the equi-stress structures by Vigdergauz \cite{Vigdergauz2018}, but is formulated globally, making de-homogenization in post-processing straightforward.
    
    The structure of this article is as follows. First, the design fields and regularization scheme are presented in \cref{sec:method:optProb:designFields}. Second, the new microstructure formulation is presented in \cref{sec:method:microGeometry}. Third, the effective properties of the multiscale material of the microstructure, including the yield strength, is calculated in \cref{sec:method:MaterialProb}. Next, the yield strength is used to formulate a multiscale stress-based Yield Load Factor (YLF) which can be used to formulate either a constraint or objective based on the macroscale stress conditions in \cref{sec:method:optProb:yieldConstraint}. Then, the simple de-homogenization procedure is presented in \cref{sec:method:dehom}. Finally, the method is demonstrated on an L-beam example in \cref{sec:results} using both arbitrary as well as real physical materials, before a summary and conclusion are given in \cref{sec:clonslusion}.

    \section{Method}
    \label{sec:method}
    This section outlines the approach for incorporating yield stress limits in multiscale material optimization. A new triangular microstructure representation is introduced, offering near-optimal isotropic stiffness and stress performance, with a two-parameter model that enables smooth globally varying de-homogenization by spatially adjusting hole size and shape. Homogenization theory is applied to determine and optimize the material's effective properties, multiscale yield limits, and buckling stability. Furthermore, a multiscale YLF is formulated based on macroscale stress conditions. Finally, a simple de-homogenization procedure is presented.
    
    To perform analysis and optimizations in this work, we use Finite Element Analysis (FEA). All analyses utilize the linear buckling assumption, which is a common approach in topology optimization \cite{Chin2016,ferrari2019a,ferrari2020a,ferrari2021a,Xu2023}. This is done by first solving a linear elastic problem
    \begin{equation} \label{eq:lineaElastic}
        \mathbf{K}(\bm{\rho}) \mathbf{u}_0 = \mathbf{f}_0,
    \end{equation}
    where $\mathbf{u}_0$ are the reference displacements corresponding to the reference load $\mathbf{f}_0$ using the density dependent stiffness matrix $\mathbf{K}({\bm{\rho}})$. The design variables are the elemental relative densities $\bm{\rho}$. The reference displacements are used to calculate the buckling load factor $\lambda$ by solving the eigenvalue problem
    \begin{equation}
        \left( \mathbf{K}(\bm{\rho}) + \lambda \mathbf{K}_{\sigma}(\bm{\rho},\mathbf{u}_0) \right) \bm{\varphi} = \mathbf{0}, \ \bm{\varphi} \neq \mathbf{0},
    \end{equation}
    where $\mathbf{K}_{\sigma}(\bm{\rho},\mathbf{u}_0)$ is the stress stiffness matrix  and $\bm{\varphi}$ are the eigenmodes. For numerical purposes, the buckling load factor is substituted with $\lambda = 1/\gamma$ during calculations. For more details on buckling topology optimization the reader is referred to the review by Ferrari et al. \cite{ferrari2021a} and Christensen et al. \cite{Christensen2023}.

    \subsection{Design Fields and Regularization}\label{sec:method:optProb:designFields}
    
    In this work, we aim to utilize isotropic multiscale material while maintaining control over the minimum and maximum porosity in the microstructure. This is crucial because the density-dependent stress limits approach zero as the relative density $\rho_e$ approaches zero, as shown in \cite{Christensen2023}. This can be observed by examining the stiffness interpolation, illustrated by the Hashin-Shtrikman upper bound \cite{Hashin1963}:
    \begin{equation} \label{eq:HSBound}
        E(\rho_e) = \dfrac{\rho_e E_0}{3-2 \rho_e},
    \end{equation}
    where $E_0$ is the Young's modulus of the base material. The local buckling stress limit $\sigma_B$ from \cite{Christensen2023,Gibson1997} is given by
    \begin{equation} \label{eq:localBucklingLimit}
        \sigma_B(\rho_e) = E_0 (b_0 \rho_e ^{n_0} + b_1 \rho_e ^{n_0+1}),
    \end{equation}
    where $b_0$ and $b_1$ are microstructure-specific constants, and $n_0$ is the power law exponent. A common feature of both \cref{eq:HSBound} and \cref{eq:localBucklingLimit} is that the limit approaches zero as the relative density $\rho_e$ approaches zero.
    
\begin{figure*}[tbp]
        \centering
        \graphicspath{{figures/multiField/}}
        \makebox[\textwidth][c]{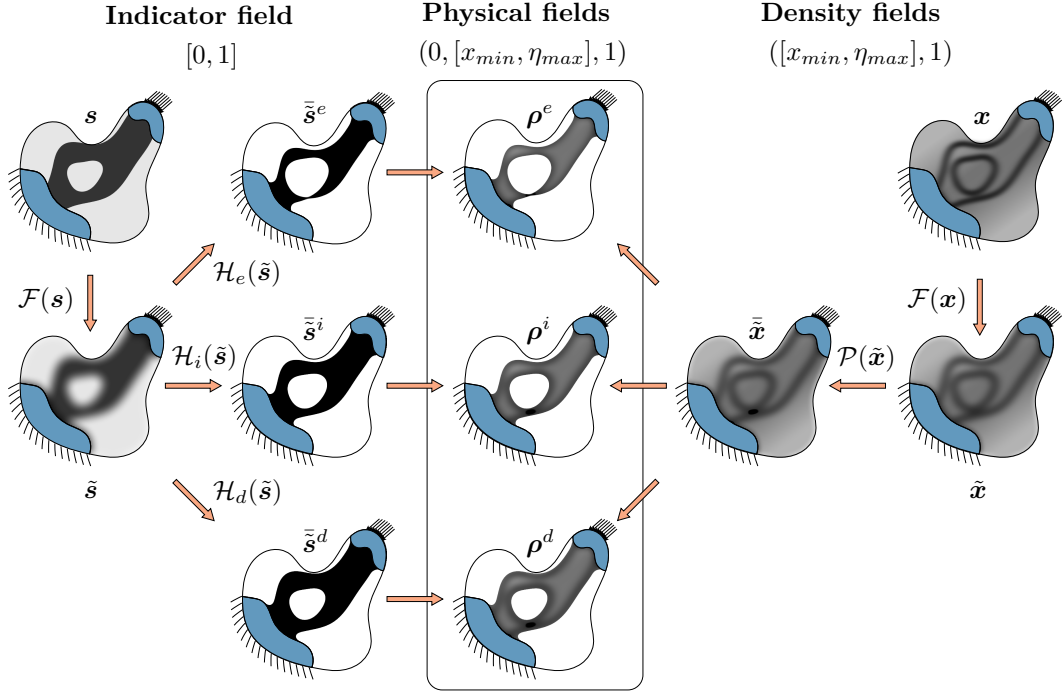}%
        \caption{Illustration of the multifield method used to create the physical design field $\bm{\rho}^m$ from the density field $\bm{x}$ and the void indicator field $\bm{s}$. The illustration is made on an arbitrary design field.}
        \label{fig:multiFieldYield}
    \end{figure*}

    Without a strategy to manage this issue, optimization would invariably require some material in every element to satisfy the constraint, effectively preventing topology changes and limiting the method to infill optimization of pre-existing geometries. To this end, we need to control the material distribution within the design domain. The physical design field $\bm{\rho}^m = \bar{\tilde{\bm{s}}}^m \bar{\tilde{\bm{x}}}$ represents the material distribution and must be either 0 (void), within the interval $\interval{x_{min}}{\eta_{max}}$ (intermediate density), or 1 (solid). Here $x_{min}$ is the minimum value on the density field $\bm{x}$, and $\eta_{max}$ is the threshold value at which $\bm{x}$ is thresholded to 1. The physical design field is described by the robust formulation where $m \in \{e,i,d\}$ represents the \textit{eroded}, \textit{intermediate}, and \textit{dilated} design fields following the multifield method introduced by Giele et al. \cite{Giele2021}. Here, $\bar{\tilde{\bm{s}}}^m$ is the void indicator field with values in the range $[0,1]$, and $\bar{\tilde{\bm{x}}}$ is the density field with values in the range $\interval{x_{min}}{\eta_{max}}$ or~1. The process for creating the physical design field $\bm{\rho}^m$ is illustrated in \cref{fig:multiFieldYield}. With this two variable formulation we can resolve the stress singularity issue by only enforcing the stress constraint on the $x$-field.
    
    To employ this we start with the design variable fields $\bm{s}$ and $\bm{x}$. The void indicator field $\bm{s}$ is filtered through $\mathcal{F}(\bm{s})$ and then projected using a Heaviside function $\mathcal{H}_m(\tilde{\bm{s}})$ at three different threshold levels. Here, $s_e = 1$ indicates the presence of material in an element, while $s_e = 0$ signifies the absence of material or a void. 
    
    Meanwhile, the density field is filtered through $\mathcal{F}(\bm{x})$. 
    Furthermore, an additional regularization scheme is applied to the density field $\tilde{\bm{x}}$ to ensure that the density field is in the region $\interval{x_{min}}{\eta_{max}}$ or 1. The purpose of this will be further elaborated in \cref{sec:method:MaterialProb:yield}. This scheme is a modified version of the one presented by Groen and Sigmund \cite{Groen2018}, as it only pushes values above a certain threshold to 1. Combining these two fields forms the physical design field $\bm{\rho}^m$.
    
    By utilizing the multifield method, stress limits are interpolated using the infill density field $\bar{\tilde{\bm{x}}}$, whereas the stiffness interpolation is applied to the physical design field $\bm{\rho}^m$.

    A density filter is included to eliminate numerical artifacts such as checkerboarding and to ensure the design is mesh-independent, as detailed in \cite{Bourdin2001}. Here a Helmholtz-type PDE filter \cite{Lazarov2011} with Robin boundary condition \cite{Wallin2020a} is used
    \begin{equation}\label{eq:densityFilter}
        \begin{split}
            \tilde{y}_e =\ &  \mathcal{F}(\bm{y}) \\ 
            \Rightarrow&  -r^2 \nabla^2 \tilde{y} + \tilde{y} = y
            , \quad \bm{y} \in \{\bm{x},\bm{s}\}, \\
            & \ r^2\nabla \tilde{y} \cdot \bm{n} = -r_\Gamma \tilde{y}, \ \text{on} \ \partial \Omega, \quad \bm{y} \in \{\bm{x},\bm{s}\},
        \end{split}
    \end{equation}
    where $r$ is the filter radius controlling the length scale. It is related to the classical filter radius for the density filters $R$ by $r = R/(2 \sqrt{3})$ \cite{Andreassen2011}, and $r_\Gamma$ is the (surface) length scale parameter. In the remainder of this work $r_\Gamma = 100 r$ to ensure room for all robust projections inside the design domain. The filter is applied to the density field $\bm{x}$ and the void indicator field $\bm{s}$, resulting in the filtered fields $\tilde{\bm{x}}$ and $\tilde{\bm{s}}$.
    
    The projection applied to the indicator field is done using the smoothed Heaviside function \cite{Wang2011, Lazarov2016}
    \begin{equation}\label{eq:Heaviside}
      \begin{split}
        \bar{\tilde{\bm{s}}} &= \mathcal{H}(\tilde{\bm{s}})\\ &= 
        \dfrac{\tanh(\beta \eta^m) + \tanh(\beta (\tilde{\bm{s}}-\eta^m))}{\tanh(\beta \eta^m)+ \tanh(\beta (\bm{1}-\eta^m))}
        , \quad m \in \{e,i,d\},
      \end{split}
    \end{equation}
    where $\beta$ is the sharpness parameter and the three robust realizations of the density field are controlled through the threshold parameter $\eta^m \in [0, 1]$. 
    
    The regularization scheme applied to the density field $\tilde{\bm{x}}$ is defined as:
    \begin{equation}\label{eq:xBarTilde}
      \begin{split}
        \bar{\tilde{\bm{x}}} =& \ \mathcal{P}(\tilde{\bm{x}}, \eta_{max}, \beta) \\
        =& \ \tilde{\bm{x}} (1 - \mathcal{H}(\tilde{\bm{x}}, \eta_{max}, \beta)) \\
        & \ + \left(\dfrac{\beta - 1}{\beta} + \frac{\tilde{\bm{x}}}{\beta}\right) \mathcal{H}(\tilde{\bm{x}}, \eta_{max}, \beta)
      \end{split}
    \end{equation}
    where $\mathcal{H}(\tilde{\bm{x}}, \eta_{max}, \beta)$ is the Heaviside function from \cref{eq:Heaviside}. A threshold of $\eta_{max} = 0.9$ is used, and the sharpness parameter $\beta$ is the same as used in the Heaviside function applied to the void indicator field $\tilde{\bm{s}}$. 

    \subsection{Two-parameter microstructure geometry}
    \label{sec:method:microGeometry}
    The novel near-optimal microstructure description used in this work is a product of two continuous fields that form one single continuous field when added together. Taking the contour at specific thresholds of this field describes the geometry of the microstructure. The way the two initial fields are handled results in different material properties. Contrary to the structures presented by Vigdergauz \cite{Vigdergauz2018}, or the formulations discussed by Norato \cite{Norato2018}, the microstructure is formulated globally periodic inspired by the formulation by Maldovan and Thomas \cite{Maldovan2008}. The first of the two fields defines the overall shape of the geometry by
    %
    \begin{multline}\label{eq:Fgeom}
        f_{geom}(y_1,y_2) = \frac{1}{3} \Bigg| \sin \left( \frac{ \pi \sqrt{3} }{d} \left( -y_1 + \frac{y_2}{\sqrt{3}} \right) \right) \\
         + \sin \left( \frac{2 \pi }{d} y_2 \right)
        + \sin \left( \frac{ \pi \sqrt{3} }{d} \left( -y_1 - \frac{y_2}{\sqrt{3}} \right) \right) \Bigg|,
    \end{multline}
    %
    %
    where $y_1$ and $y_2$ are the spatial coordinates and $d$ controls the size (lattice parameter) of the microstructure by defining the distance between two parallel lines in the microstructure, as depicted in \cref{fig:multiscaleIllustration}.
    
    The second field is used to control the sharpness of the corners in the microstructures geometry. It is defined by
    \begin{multline}\label{eq:Fsharp}
        f_{sharp}(y_1,y_2) = \\\frac{1}{3} \Bigg( 
            \cos \left( \frac{\pi \sqrt{3}}{d} 
            \left( -y_1 - \frac{d \sqrt{3}}{6} + \frac{1}{\sqrt{3}} \left(y_2 + \frac{d}{2}\right) \right) \right) \\
            + \cos \left( \frac{2 \pi}{d} \left(y_2 + \frac{d}{2}\right) \right) \\
            + \cos \left( \frac{\pi \sqrt{3}}{d} \left( -y_1 - \frac{d \sqrt{3}}{6} - \frac{1}{\sqrt{3}} \left(y_2 + \frac{d}{2}\right) \right) \right) 
        \Bigg).
    \end{multline}
    The final geometry is defined as
    \begin{equation}\label{eq:Ftriangle}
        f_{triangle}(y_1,y_2,\alpha) = f_{geom}(y_1,y_2) + \alpha f_{sharp}(y_1,y_2),
    \end{equation}
    where $\alpha$ is a variable controlling the sharpness of the corners. Finally, the geometry is determined using a threshold $\bar{\eta}$ to threshold \cref{eq:Ftriangle} to either 0 and 1, using 
    \begin{equation} \label{eq:dehom:heaviside}
        {\bm{\rho}}_Y = H(f_{triangle}(y_1,y_2,\alpha),\bar{\eta}),
    \end{equation}
    or by extracting the contour at the threshold,
    \begin{equation} \label{eq:dehom:contour}
        \Gamma_Y = \mathcal{C}(f_{triangle}(y_1,y_2,\alpha),\bar{\eta}).
    \end{equation}
    By taking the contour, the boundaries of the geometry can be imported into commercial software such as COMSOL, enabling body-fitted meshes to be used. 
    
\begin{figure*}
        \centering
        \graphicspath{{figures/}}
        \makebox[\textwidth][c]{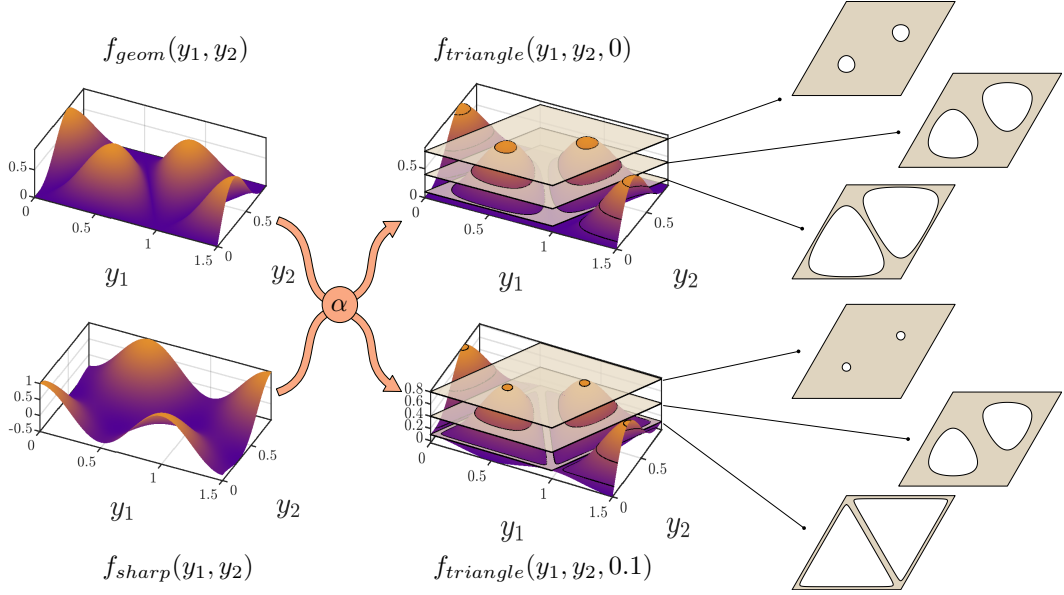}%
        \caption{Illustration of the two-parameter microstructure defined by $f_{geom}$ and $f_{sharp}$ with two values of $\alpha \in [0,0.1]$ and three values of $\bar{\eta} \in[0,0.4,0.8]$.}
        \label{fig:twoTermGeometry}
    \end{figure*}

    An illustration of the two-parameter microstructure is presented in \cref{fig:twoTermGeometry} with different values of the two parameters $\alpha$ and $\bar{\eta}$. As visible in the figure, the values of the two parameters influence the geometry of the microstructure. This, in turn, results in different macroscale material properties and densities, meaning that the material properties can be tuned by the two parameters. To investigate the parameters influence on the volume, stiffness and yield limits, a parameter sweep is performed over $\alpha = \interval{-0.65}{0.22}$ and $\bar{\eta} = \interval{0.005}{1.25}$. 
    
\begin{figure*}[tb]
        \centering
        \makebox[\textwidth][c]{
        \begin{subfigure}[b]{0.4\textwidth}
                \vspace{0.5cm}
    \begin{tikzpicture}
        \begin{axis}[
            xlabel = {$\bar{\sigma}_1$},
            ylabel = {$\bar{\sigma}_2$},
            xmin=-1.2, xmax=1.2,
            ymin=-1.2, ymax=1.2,
            width=7cm,   
            height=7cm,    
            xmajorgrids=true,
            ymajorgrids=true,
        ]
        
        \addplot[
            color=mycolor1,
            line width=1pt,
            mark=*,
        ] coordinates {(-1,1) (-1,-1)};
        \addplot[
            color=mycolor1,
            -{Latex[length=3mm]}
        ] coordinates {(-1,0.35) (-1,-0.15)};
        
        \addplot[
            color=mycolor1,
            line width=1pt,
            mark=*,
        ] coordinates {(-1,-1) (1,-1)};
        \addplot[
            color=mycolor1,
            -{Latex[length=3mm]}
        ] coordinates {(-0.35,-1) (0.15,-1)};
        
        \addplot[
            color=mycolor1,
            line width=1pt,
            mark=*,
        ] coordinates {(1,-1) (1,1)};
        \addplot[
            color=mycolor1,
            -{Latex[length=3mm]}
        ] coordinates {(1,-0.35) (1,0.15)};
        
        \addplot[
            color=mycolor1,
            line width=1pt,
            mark=*,
        ] coordinates {(1,1) (-1,1)};
        \addplot[
            color=mycolor1,
            -{Latex[length=3mm]}
        ] coordinates {(0.35,1) (-0.15,1)};
        
        \end{axis}
    \end{tikzpicture}
        \caption{}
            \label{fig:yieldHomogenization:1}
        \end{subfigure}
        \hfill
        \begin{subfigure}[b]{0.55\textwidth}
        \input{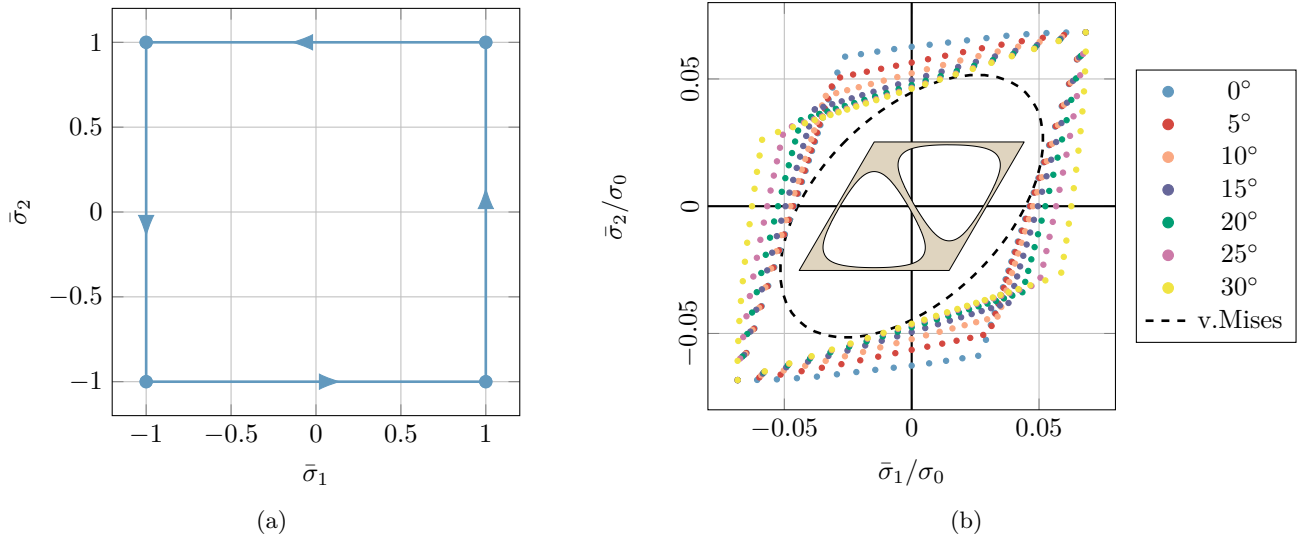}
    
        \caption{}
            \label{fig:yieldHomogenization:2}
        \end{subfigure}
        }
        \caption{Illustration of the yield strength analyses for the two-parameter microstructure. (a) Illustration of the zone of evaluated macroscopic stress states. (b) Microstructural yield strength for all principal stress states and cell orientations of the microstructure geometry in the center. The dashed line indicates the inscribed von Mises yield surface.}
        \label{fig:yieldHomogenization}
    \end{figure*}

    The macroscale material properties, such as bulk $\kappa$ and shear $\mu$ modulus, are determined using homogenization theory \cite{Guedes1990a, sigmund1994a}. The yield strength is determined following Andersen et al. \cite{Andersen2021a,Andersen2022}. First the microscale stresses are found using
    \begin{equation}
        \bm{\sigma}_e = \mathbf{C}_e (\mathbf{I}-\mathbf{B}_e \mathbf{X}_e) \bar{\bm{\varepsilon}}
    \end{equation}
    where $\mathbf{C}_e$ is the microscale constitutive matrix for the base material. $\mathbf{I}$ is the identity matrix, $\mathbf{B}_e$ is the strain-displacement matrix. The element test field obtained from homogenization are arranged in $\mathbf{X}_e = [\chi_e^1,\chi_e^2,\chi_e^3]$ and $\bar{\bm{\varepsilon}}$ is the homogenized macroscale strain  obtained from the macroscale stress $\bar{\bm{\sigma}}$ using the homogenized constitutive relation
    \begin{equation}
        \bar{\bm{\varepsilon}} = \mathbf{C}_e^{H^{-1}} \bar{\bm{\sigma}}
    \end{equation}
    where $\mathbf{C}_e^{H}$ is the homogenized constitutive matrix. The macroscale stresses are evaluated at states ranging from $\bar{\bm{\sigma}} = [-1,1,0]$ via $\bar{\bm{\sigma}} = [-1,-1,0]$, $\bar{\bm{\sigma}} = [1,-1,0]$, $\bar{\bm{\sigma}} = [1,1,0]$ and back to $\bar{\bm{\sigma}} = [-1,1,0]$, (see \cref{fig:yieldHomogenization:1}). Furthermore, the unit cell is evaluated at seven different orientations in the interval $\theta = \interval{\SI{0}{\degree}}{\SI{30}{\degree}}$ to capture the worst-case scenario. The maximum von Mises stress in the unit cell $\sigma_{vm,max}$ is calculated at each state and used to determine the yield strength $\sigma_y$ of the multiscale material. This is done through the relation
    \begin{equation}
        \sigma_y = \tilde{\sigma}_y \sigma_0 = \min_{\bar{\bm{\sigma}},\theta}\left(\frac{\bar{\sigma}_{vm}(\bar{\bm{\sigma}},\theta)}{\sigma_{vm,max}(\bar{\bm{\sigma}},\theta)}\right) \sigma_0,
    \end{equation} 
    where $\tilde{\sigma}_y$ is the local yield strength factor of the multiscale material relating the base material yield strength $\sigma_0$ to the microscale yield strength $\sigma_y$. The local yield strength factor is determined by the minimum ratio between $\sigma_{vm,max}$ and the macroscale von Mises stress $\bar{\sigma}_{vm}$ to generate a conservative inscribed von Mises yield surface as illustrated in \cref{fig:yieldHomogenization:2} for the arbitrary parameters $\alpha=-0.02$ and $\bar{\eta}= 0.1$. For this configuration the yield strength factor is $\tilde{\sigma}_y = 0.0447$.
    
\begin{figure*}
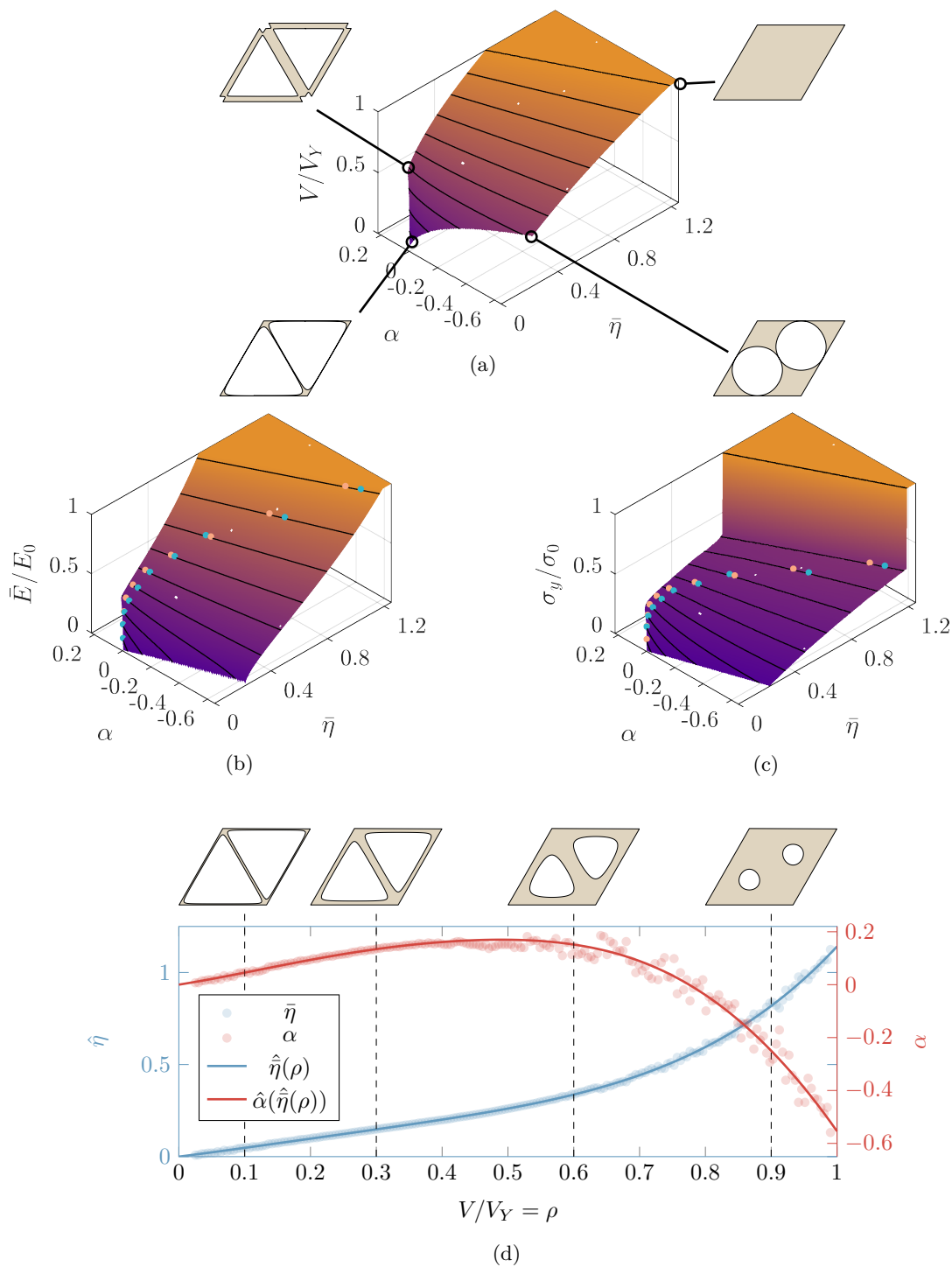

        \centering        
            \begin{subfigure}[]{0.79\textwidth}
                \makebox[\textwidth][c]{
                    \input{figures/YieldParmaterFit/maxSigmaParameterFit.tex}
                }
            \end{subfigure}
            \begin{subfigure}[]{0.79\textwidth}
                \vspace{0.5cm}
                \makebox[\textwidth][c]{
                    \input{figures/YieldParmaterFit/maxSigmaParameterFit2.tex}
                }
                \caption{}
                \label{fig:parameterSweep:4}
                
            \end{subfigure}
        \caption{Results of the parameter sweep over $\alpha$ and $\bar{\eta}$. (a) Volume fraction with isocontours indicating constant volume paths. (b) Effective relative Young's modulus with volume isocontours projected onto the stiffness surface. Peach colored dots mark the maximum Young's modulus and blue dots the maximum yield strength along each isocontour. (c) Relative yield strength with volume isocontours projected onto the surface. (d) Parameters $\alpha$ and $\bar{\eta}$ relative to the volume fraction for the yield strength optimal microstructures.}
        \label{fig:parameterSweep}
    \end{figure*}

    
    The microstructure volume fraction $V/V_Y$, effective relative Young's modulus $\bar{E}/E_0$, and relative yield strength $\sigma_y/\sigma_0$ obtained from the parameter sweep are shown in \cref{fig:parameterSweep}. \cref{fig:parameterSweep:1} presents the volume fraction with isocontours indicating lines of constant volume. These isocontours are projected onto the effective relative stiffness in \cref{fig:parameterSweep:2}. The stiffness along each isocontour is evaluated, and a peach colored dot indicates the maximum value. For visualization purposes, only ten isocontours are shown, but 200 are evaluated as visible in \cref{fig:parameterSweep:4}. For low volume fractions ($V/V_Y<0.5$), increasing $\alpha$, i.e., sharper corners, produces structures with higher stiffness. By having sharp corners, the material is prioritized in the bars to make them as thick as possible, thus significantly increasing the area moment of inertia. For structures with $V/V_Y > 0.5$, $\alpha$ decreases and becomes negative, resulting in rounder holes. By doing this, the effective thickness-to-length ratio is increased, resulting in higher stiffness.
    
    \cref{fig:parameterSweep:3} shows the microstructural relative yield strength. A noticeable jump in yield strength occurs when the microstructure transitions to a fully solid configuration. This step brings the yield strength down to $\sigma_y = 0.327\sigma_0$, indicating that introducing holes in the microstructure decreases the yield strength to approximately one-third of the base material's yield strength. Comparing this to Kirsch equation \cite{Kirsch1898}, which gives $\sigma_y = 0.327\sigma_0$, the results align well.

    The blue dots in \cref{fig:parameterSweep:3} mark the maximum yield strength along each isocontour. The trend with the initially increasing, then decreasing value of $\alpha$ observed in \cref{fig:parameterSweep:2} is repeated for the yield strength. Comparing the optimal stiffness and yield strength parameters only show minor differences. The fluctuation in the maximum yield limit $\sigma_y$, observed for the higher volume fractions, results from discretization errors in the FEA. 
    
    Selecting the parameters to describe the geometry of the microstructure can be based on which objective is most important. The data in \cref{fig:parameterSweep} shows maximum stiffness and yield strength. Achieving maximum buckling stability is also possible but comes at a significantly higher computational price \cite{thomsen2018a}. However, the added rounding at the corners most likely also has a positive effect on the buckling stability as small fillets can increase the thickness-to-length ratio enough to improve buckling stability as well \cite{Hubner2023}. This section presents the maximum yield strength parameters. The parameters required to achieve maximum stiffness are presented in Appendix A.

    The parameters are linked to the relative volume of the microstructure through curve fitting. The relative volume fraction directly relates to the relative densities $\rho_e$ and thresholds $\bar{\eta}$ are fitted relative to the volumes in \cref{fig:parameterSweep:1} using a \nth{5} order polynomial.
    \begin{equation}\label{eq:etaBarFit}
        \hat{\bar{\eta}}(\rho_e) = p_1 \rho_e^5 + p_2 \rho_e^4 + p_3 \rho_e^3 + p_4 \rho_e^2 + p_5 \rho_e ,
    \end{equation}
    where the coefficients $p_1$--$p_5$ are defined in \cref{tab:etaBarFit}. The factor $\alpha$ is fitted to the thresholds $\bar{\eta}$ using a rational function
    %
        \begin{center}
        \captionof{table}{Values of the fitted coefficients used in $\hat{\bar{\eta}}(\rho_e)$.}\label{tab:etaBarFit}
        \begin{tabular}{rrrrrr}
         \midrule
         $\bm{p_1}$ & $\bm{p_2}$ & $\bm{p_3}$ & $\bm{p_4}$  & $\bm{p_5}$  \\
         0.3781 & 0.8792 &  -0.8795 & 0.3104 & 0.4532  \\
         \bottomrule
         \end{tabular}
        \end{center}
    %
    \begin{equation}\label{eq:alphaFit}
        \hat{\alpha}(\hat{\bar{\eta}}(\rho_e)) = \dfrac{r_1 \hat{\bar{\eta}}(\rho_e)^3 + r_2 \hat{\bar{\eta}}(\rho_e)^2 + r_3 \hat{\bar{\eta}}(\rho_e) + r_4}
        {\hat{\bar{\eta}}(\rho_e)^3 + q_1 \hat{\bar{\eta}}(\rho_e)^2 + q_2 \hat{\bar{\eta}}(\rho_e) + q_3}
    \end{equation}
    The coefficients $r_1$--$r_4$ and $q_1$--$q_3$ are listed in \cref{tab:alphaFit}. The fits for \cref{eq:etaBarFit} and \cref{eq:alphaFit} are shown in \cref{fig:parameterSweep:4}. The figure also illustrates the microstructure at four different relative densities. For low densities, the microstructure features thin bars with sharp corners, which become rounder as the density increases, resembling the equi-stress-shaped Vigdergauz structures \cite{Vigdergauz2018}.
    %
    \begin{center}
        \captionof{table}{Values of the fitted coefficients used in $\hat{\alpha}(\rho_e)$.}\label{tab:alphaFit}
        \begin{tabular}{rrrrrrr}
         \midrule
         $\bm{r_1}$ & $\bm{r_2}$ & $\bm{r_3}$ & $\bm{r_4}$ & $\bm{q_1}$ & $\bm{q_2}$ & $\bm{q_3}$ \\
         -7164 & 2106 & 1017  & 0.059 & 8770 & -354.7 & 1153 \\
         \bottomrule
         \end{tabular}
    \end{center}
    %

    \subsection{Homogenized Material Properties}\label{sec:method:MaterialProb}
    The effective material properties of the microstructure defined in \cref{sec:method:microGeometry} are determined offline, prior to any optimization. The microstructure is evaluated at thirty relative densities in the interval $\rho \in \{0.05,1\}$ using the parameter interpolations from \cref{eq:etaBarFit} and \cref{eq:alphaFit}. In addition to stiffness and yield strength, buckling strength is also evaluated for these densities. Yield and stiffness properties are determined following the procedure in \cref{sec:method:microGeometry}. The buckling limits are determined using Bloch-Floquet cell analysis following the approaches described in \cite{Triantafyllidis1998,thomsen2018a,Christensen2023}.
    
    \subsubsection{Homogenized Stiffness}\label{sec:method:MaterialProb:stiffness}
    The homogenized material properties are interpolated using a Rational Approximation of Material Properties (RAMP) method for both bulk $\kappa$ and shear $\mu$ moduli. The RAMP functions are defined as follows:
    \begin{align}
        \hat{\mu}(\rho_e^m) &= \mu_{min} + \frac{\rho_e^m}{1 + q_{\mu} (1 - \rho_e^m)} (\mu_0 - \mu_{min}), \label{eq:shearFit}\\
        \hat{\kappa}(\rho_e^m) &= \kappa_{min} + \frac{\rho_e^m}{1 + q_{\kappa} (1 - \rho_e^m)} (\kappa_0 - \kappa_{min}), \label{eq:bulkFit}
    \end{align}
    where $\mu_{min}$ and $\kappa_{min}$ are the minimum shear and bulk moduli, respectively, and $\mu_0$ and $\kappa_0$ are the moduli of the base material. The parameters $q_{\mu}$ and $q_{\kappa}$ are used to fit the interpolation functions to the homogenized data. The fitted parameters are listed in \cref{tab:shearBulkFit}, and the corresponding fits are shown in \cref{fig:shearBulkFit}. The figure also includes the Hashin-Shtrikman (HS) upper bounds for the bulk and shear moduli, demonstrating the near-optimal isotropic stiffness of the microstructure ($q_{\mu,HS} = q_{\kappa,HS} = 2$).
    \begin{Table}
        \centering
        \captionof{table}{Values of the fitted coefficients used to interpolate bulk and shear modulus}\label{tab:shearBulkFit}
        \begin{tabular}{rr}
         \midrule
         $\bm{q_{\mu}}$ & $\bm{q_{\kappa}}$ \\
         2.097 & 2.0  \\
         \bottomrule
         \end{tabular}
    \end{Table}
    %
    \begin{Figure}
        \centering
                \begin{tikzpicture}
\pgfplotsset{
    scale only axis,
    xmin=0, xmax=1
}

    \pgfmathsetmacro{\qMu}{2.097} 
    \pgfmathsetmacro{\qKappa}{2.0} 

    \pgfmathsetmacro{\k}{0.75} 
    \pgfmathsetmacro{\kMin}{0.0} 

    \pgfmathsetmacro{\muZero}{0.375} 
    \pgfmathsetmacro{\muMin}{0.0} 

    \pgfmathdeclarefunction{muFit}{1}{%
        \pgfmathparse{\muMin + #1/(1+\qMu*(1-#1))*(\muZero-\muMin)}%
        \let\pgfmathresult=\pgfmathresult
    }

    \pgfmathdeclarefunction{kappaFit}{1}{%
        \pgfmathparse{\kMin + #1/(1+\qKappa*(1-#1))*(\k-\kMin)}%
        \let\pgfmathresult=\pgfmathresult
    }

    \pgfmathdeclarefunction{kappaHS}{1}{%
        \pgfmathparse{#1*\k+(1-#1)*\kMin-#1*(1-#1)*(\k-\kMin)^2/((1-#1)*\k+#1*\kMin+\muZero)}%
        \let\pgfmathresult=\pgfmathresult
    }

    \pgfmathdeclarefunction{muHS}{1}{%
        \pgfmathparse{#1*\muZero+(1-#1)*\muMin-#1*(1-#1)*(\muZero-\muMin)^2/((1-#1)*\muZero+#1*\muMin+\k*\muZero/(\k+2*\muZero))}%
        \let\pgfmathresult=\pgfmathresult
    }



    \begin{axis}[
        xlabel = {$\rho$},
        ylabel = {$\mu(\rho), \ \kappa(\rho)$},
        domain=0:1,
        legend pos = north west,
        ymin=0, ymax=0.8,
        width=5.5cm,   
        height=5.5cm,    
        grid=both,
    ]
    \addplot+[
        only marks,
        mark=*,
        mark size=2.5pt,
        color = mycolor1,
        fill=mycolor1,
        mark options={solid},
        inner sep=0.5pt,
        fill opacity=0.4,
        draw opacity=0,
     ]
    table [x index=0, y index=2, col sep=comma] {figures/bulkShearFit/bulkShearData.csv};
    \addlegendentry{$\mu$}

    \addplot+[
        only marks,
        mark=*,
        mark size=2.5pt,
        color = mycolor2,
        fill=mycolor2,
        mark options={solid},
        inner sep=0.5pt,
        fill opacity=0.3,
        draw opacity=0,
     ]
    table [x index=0, y index=1, col sep=comma] {figures/bulkShearFit/bulkShearData.csv};
    \addlegendentry{$\kappa$}

    \addplot[mycolor1,
            line width = 1.5,
            domain=0:1,
            samples=100]
            {muFit(x)};
            \addlegendentry{$\hat{\mu}(\rho)$}
    
    \addplot[mycolor2,
            line width = 1.5,
            domain=0:1,
            samples=100]
            {kappaFit(x)};
            \addlegendentry{$\hat{\kappa}(\rho)$}

    \addplot[black,
            dotted,
            line width = 1.5,
            domain=0:1,
            samples=100]
            {muHS(x)};
            \addlegendentry{$\mu_{HS}(\rho)$}

    \addplot[black,
            loosely dashed,
            line width = 1.5,
            domain=0:1,
            samples=100]
            {kappaHS(x)};
            \addlegendentry{$\kappa_{HS}(\rho)$}
\end{axis}




\end{tikzpicture}%
        \captionof{figure}{Bulk and shear values with fitted interpolation curves and the HS upper bounds for isotropic material.}
        \label{fig:shearBulkFit}
    \end{Figure}

    \subsubsection{Homogenized Yield Strength}\label{sec:method:MaterialProb:yield}
    The density-dependent yield limit is defined using an extended interpolation scheme inspired by Andersen et al. \cite{Andersen2021a}:
    \begin{equation}\label{eq:yieldLimit}
        \bar{\sigma}_{lim}(\bar{\tilde{\bm{x}}}) = \sigma_{0} \left(c_0 \bar{\tilde{\bm{x}}} + c_1 \bar{\tilde{\bm{x}}}^2 + c_2 \bar{\tilde{\bm{x}}}^3 + c_3 \bar{\tilde{\bm{x}}}^4 + c_4 \bar{\tilde{\bm{x}}}^5\right),
    \end{equation}
    where $c_0$--$c_4$ are the fitted coefficients listed in \cref{tab:yieldFitCoefficients}. 
    
    To account for the jump in the stress limit at the transition from having holes to not having holes in the microstructure, the yield stress limit is relaxed using a Heaviside function. This provides a smooth representation of the yield stress limit jump:
    \begin{equation}\label{eq:relaxedYieldLimit}
        \sigma_{lim}(\bar{\tilde{\bm{x}}}) = 
        \bar{\sigma}_{lim}(\bar{\tilde{\bm{x}}}) + \left(  \sigma_{0} - \bar{\sigma}_{lim}(1) \right) \mathcal{H}(\bar{\tilde{\bm{x}}},\eta_{yield},\beta_{yield}),
    \end{equation}
    where $\eta_{yield} = 0.999$ is the threshold representing the yield stress limit jump. The sharpness parameter $\beta_{yield}$ is controlled through a continuation scheme starting from $\beta_{yield} = 2$ and ending at $\beta_{yield}= 50$. The relaxed interpolation in \cref{eq:relaxedYieldLimit} is illustrated in \cref{fig:twoTermRelaxations} with the yield limit data from the homogenized analysis and the infill thresholding defined by \cref{eq:xBarTilde}. The projection ensures that the optimizer does not exploit the yield limit jump by making sure that densities are in the region $\interval{0}{\eta_{max}}$ or 1.
    \begin{Table}
        \centering
        \captionof{table}{Values of the fitted coefficients used for the yield limit interpolation of the microstructure from \cref{sec:method:microGeometry}.}\label{tab:yieldFitCoefficients}
        \begin{tabular}{rrrrr}
         \multicolumn{5}{c}{Yield Limit Coefficients} \\
         \midrule
         $\bm{c_0}$ & $\bm{c_1}$ & $\bm{c_2}$ & $\bm{c_3}$ & $\bm{c_4}$ \\
         0.2266 & 0.8066 & -2.5827 & 2.9753 & -1.0987 \\
         \bottomrule
         \end{tabular}
    \end{Table}
    %
    \begin{Figure}
        \centering
            \begin{tikzpicture}    
    \pgfmathsetmacro{\czero}{0.2266} 
    \pgfmathsetmacro{\cone}{0.8066}  
    \pgfmathsetmacro{\ctwo}{-2.5827} 
    \pgfmathsetmacro{\cthree}{2.9753} 
    \pgfmathsetmacro{\cfour}{-1.0987} 
    \pgfmathsetmacro{\sigmazero}{1}  
    \pgfmathsetmacro{\eta}{0.999} 
    \pgfmathsetmacro{\beta}{50} 
    \pgfmathsetmacro{\etamax}{0.9} 
    \pgfmathsetmacro{\betatilde}{100} 
    \pgfmathdeclarefunction{f}{1}{%
        \pgfmathparse{\sigmazero * (\czero * #1 + \cone * #1^2 + \ctwo * #1^3 + \cthree * #1^4 + \cfour * #1^5)}%
        \let\pgfmathresult=\pgfmathresult
    }
    \pgfmathdeclarefunction{H}{3}{%
    \pgfmathparse{
        (tanh(#3 * #2) + tanh(#3 * (#1 - #2))) / 
        (tanh(#3 * #2) + tanh(#3 * (1 - #2)))
    }%
    \let\pgfmathresult=\pgfmathresult
    }
    \pgfmathdeclarefunction{g}{1}{%
    \pgfmathparse{f(#1) + (\sigmazero - f(1)) * H(#1, \eta, \beta)}%
    \let\pgfmathresult=\pgfmathresult
    }
    \pgfmathdeclarefunction{h}{1}{%
    \pgfmathparse{#1 * (1 - H(#1, \etamax, \betatilde)) + ((\betatilde - 1)/\betatilde + #1/\betatilde) * H(#1, \etamax, \betatilde)}%
    \let\pgfmathresult=\pgfmathresult
    }
    \begin{axis}[
        xlabel={${\tilde{x}}, \ \bar{\tilde{x}}$},
        ylabel={$\bar{\tilde{x}}, \ {\sigma}_{{lim}}(\bar{\tilde{x}})/\sigma_0$},
        domain=0:1,
        legend pos = north west,
        xmin=0, xmax=1,
        ymin=0, ymax=1,
        width=7cm,   
        height=7cm,    
        grid=both,
    ]

    \addplot+[
        only marks,
        mark=*,
        mark size=2.5pt,
        color = mycolor1,
        fill=mycolor1,
        mark options={solid},
        inner sep=0.5pt,
        fill opacity=0.3,
        draw opacity=0,
     ]
    table [x index=0, y index=1, col sep=comma] {figures/twoTermRelaxations/sigmaData.csv};
    \addlegendentry{$\sigma_{y}$}

    \addplot[black,
            dashed,
            line width = 1.5,
            domain=0:1,
            samples=100]
            {h(x)};
            \addlegendentry{$\bar{\tilde{x}}(\tilde{x})$}
    \addplot[mycolor1,
            line width = 1.5,
            domain=0:1,
            samples=100]
            {f(x)};
            \addlegendentry{$\bar{\sigma}_{{lim}}(\bar{\tilde{x}})$}
    \addplot[mycolor2,
            dashed,
            line width = 1.5,
            domain=0:1,
            samples=100]
            {g(x)};
            \addlegendentry{${\sigma}_{{lim}}(\bar{\tilde{x}})$}
    \addplot[black, 
            line width = 1.,
            dotted] coordinates {(0,0) (1,1)};
    \end{axis}
\end{tikzpicture}
        \captionof{figure}{Yield stress limit data $\sigma_y$ and fitted interpolations $\bar{\sigma}_{{lim}}(\bar{\tilde{x}})$ and ${\sigma}_{{lim}}(\bar{\tilde{x}})$ with $\beta_{yield} = 50$. The projected density field $\bar{\tilde{x}}$ with $\eta_{max} = 0.9$ ensures that the optimization does not take advantage of the yield limit jump.}
        \label{fig:twoTermRelaxations}
    \end{Figure}

    \subsubsection{Homogenized Buckling Strength}\label{sec:method:MaterialProb:buckling}
    
    The buckling stability of the microstructure in \cref{sec:method:microGeometry} is determined using the method from \cite{Christensen2023}. Here the unified stress method by \cite{Giraldo-Londono2020} using the Willam-Warnke (WW) failure surface is used to describe the density dependent stress failure surface. For the uniaxial compression $c$ and equi-biaxial compression $b$ cases, the buckling limit is interpolated the functions from \cite{Andersen2021a}. The interpolation is defined as
    \begin{equation}\label{eq:AndersenBucklingInterpolation}
        \bar{\sigma}_{k}(\bar{\tilde{x}}_e) = E_0 \left(b_{0,k} \bar{\tilde{x}}_e^{n_0} + b_{1,k} \bar{\tilde{x}}_e^{n_0+1}\right), \quad k \in \{c,b\},
    \end{equation}
    where $n_0 = 3$ and $b_{0,k}$ and $b_{1,k}$ are the fitted coefficients obtained by curve fitting.
    
    To achieve a better fit of the WW failure surface, the interpolation of the stress limit in uniaxial tension $t$ is modified by adding an extra term, which yields
    \begin{equation}\label{eq:AndersenBucklingInterpolationModified}
        \bar{\sigma}_t(\bar{\tilde{x}}_e) = E_0 \left(b_{0,t} \bar{\tilde{x}}_e^{n_0} + b_{1,t} \bar{\tilde{x}}_e^{n_0+1} + b_{2,t} \bar{\tilde{x}}_e^{n_0+2}\right).
    \end{equation}
    All the fitted coefficients are presented in \cref{tab:bucklingFitCoefficients2}.
    \begin{Table}
        \centering
        \captionof{table}{Values of the fitted coefficients used for the buckling limit interpolation of the microstructure from \cref{sec:method:microGeometry}.}\label{tab:bucklingFitCoefficients2}
         \begin{tabular}{lrrr}
            \multicolumn{4}{c}{Buckling Limit Coefficients} \\
            \midrule
            $\bm{k}$   & $\bm{t}$ & $\bm{c}$ & $\bm{b}$ \\
           $b_{0,k}$ & -0.06751 & 0.05 & 0.0133 \\
           $b_{1,k}$ & 1.741 & 0.1644 & 0.1539 \\
           $b_{2,k}$ & -0.94 & - & - \\
            \bottomrule
            \end{tabular}
    \end{Table}

    \subsection{Stress-based Yield Load Factor}\label{sec:method:optProb:yieldConstraint}
    The formulation of the multiscale YLF leverages the unified approach for stress-based optimization \cite{Giraldo-Londono2020}. Using the microscale yield strength definition from \cref{sec:method:microGeometry}, the yield failure surface is defined using the von Mises failure criterion. This conservative criterion effectively uses the inscribed von Mises failure surface, as seen in \cref{fig:yieldHomogenization:2}.
    
    The yield failure criterion is formulated by calculating an equivalent stress, which for the von Mises failure criterion reduces to
    \begin{equation}\label{eq:FailureSurface}
        \sigma_{eq}(\rho_e^m,\bar{\tilde{x}}_e) = \frac{ 1}{\sigma_{lim}(\bar{\tilde{{x}}}_e)} \sqrt{3 J_2(\rho_e^m,\bar{\tilde{x}}_e)} ,
    \end{equation}
    where $J_2(\rho_e^m,\bar{\tilde{x}}_e)$ represents the second invariant of the deviatoric stress tensor $\bm{s}(\rho_e^m,\bar{\tilde{x}}_e)$ defined as
    \begin{equation}
        \bm{s}(\rho_e^m,\bar{\tilde{x}}_e) = \bm{\sigma}(\rho_e^m,\bar{\tilde{x}}_e) - \dfrac{I_1(\rho_e^m,\bar{\tilde{x}}_e)}{3}\mathbf{I}.
    \end{equation} 
    where $\mathbf{I}$ is the identity matrix and $I_1(\rho_e^m,\bar{\tilde{x}}_e)$ is the first invariant of the Cauchy stress tensor calculated as
    \begin{equation}
        \bm{\sigma}(\rho_e^m,\bar{\tilde{x}}_e) = 
        \mathbf{C}^H(\bar{\tilde{x}}_e) \mathbf{B} \mathbf{u}_0(\rho_e^m),
    \end{equation}
    where $\mathbf{C}^H(\bar{\tilde{x}}_e)$ comes from the homogenization-based material properties interpolated by \cref{eq:shearFit} and \cref{eq:bulkFit}. $\mathbf{B}$ is the strain-displacement matrix, and $\mathbf{u}_0(\rho_e^m)$ is the reference displacement from the linear-elastic state problem in \cref{eq:lineaElastic}. This formulation of the Cauchy stresses, where the density field $\bar{\tilde{x}}_e$ is used to interpolate the material properties, effectively means that stresses in void regions are unphysical but correct in regions where multiscale material is allowed by the indicator field. By using this approach, the indicator field is used to treat the singularity problem \cite{Le2010} using $\varepsilon$-relaxation of the stresses \cite{Guo1997, Duysinx1998, DaSilva2019}. The relaxation is
    \begin{equation}\label{eq:epsilonRelaxation}
        f_{\sigma}(\bar{\tilde{s}}_e) = \dfrac{\bar{\tilde{s}}_e}{\varepsilon (1-\bar{\tilde{s}}_e)+\bar{\tilde{s}}_e},
    \end{equation}
    where $\varepsilon$ is the relaxation parameter. By multiplying \cref{eq:epsilonRelaxation} with the equivalent stresses $\sigma_{eq}(\bm{\rho}^m,\bar{\tilde{\bm{x}}})$, the relaxed stresses are obtained
    \begin{equation}
        \sigma_{rel}(\bm{\rho}^m,\bar{\tilde{\bm{x}}},\bar{\tilde{\bm{s}}}^m) = f_{\sigma}(\bar{\tilde{\bm{s}}}^m) \sigma_{eq}(\bm{\rho}^m,\bar{\tilde{\bm{x}}}).
    \end{equation}
    
    All the yield stress constraint are aggregated using the $p$-norm defined as
    \begin{equation}\label{eq:pNorm}
      \begin{split}
        \max_{\forall e}(\sigma_{rel}({\rho}^m_e,\bar{\tilde{{x}}}_e,\bar{\tilde{{s}}}_e^m)) \approx& \ \left(\sum_e \sigma_{rel}({\rho}_e^m,\bar{\tilde{{x}}}_e,\bar{\tilde{{s}}}_e^m)^p \right)^{\frac{1}{p}} \\ =& \ \sigma_{PN}(\bm{\rho}^m,\bar{\tilde{\bm{x}}},\bar{\tilde{\bm{s}}}^m),
      \end{split}
    \end{equation}
    where $p \rightarrow \infty$ approaches $\max_{\forall e}(\sigma_{rel}({\rho}^m_e,\bar{\tilde{{x}}}_e,\bar{\tilde{{s}}}_e^m))$. For a better approximation of the actual max value, a correction $c$ following \cite{Le2010} is applied to $\sigma_{PN}$ to obtain the final YLF ($\lambda_{Y}$) definition
    \begin{equation}\label{eq:YLF}
        \lambda_{Y}(\bm{\rho}^m,\bar{\tilde{\bm{x}}},\bar{\tilde{\bm{s}}}^m) = 
        \dfrac{1}{c \sigma_{PN}(\bm{\rho}^m,\bar{\tilde{\bm{x}}},\bar{\tilde{\bm{s}}}^m)}.
    \end{equation}
    The YLF can be used either as an objective or as a constraint. When used as an objective, where $g_y^{obj}$ is minimized to maximize the YLF, it is defined as
    \begin{equation}\label{eq:yieldObejctive}
        g_y^{obj}(\bm{\rho}^m,\bar{\tilde{\bm{x}}},\bar{\tilde{\bm{s}}}^m) = \dfrac{1}{\lambda_{Y}(\bm{\rho}^m,\bar{\tilde{\bm{x}}},\bar{\tilde{\bm{s}}}^m)} = 
        c \sigma_{PN}(\bm{\rho}^m,\bar{\tilde{\bm{x}}},\bar{\tilde{\bm{s}}}^m).
    \end{equation}
    When used as a constraint, \cref{eq:yieldObejctive} is rearranged such that the yield stress constraint is defined as
    \begin{equation}\label{eq:yieldConstraint0}
        \begin{split}
        g_y^{const}(\bm{\rho}^m,\bar{\tilde{\bm{x}}},\bar{\tilde{\bm{s}}}^m) &=  
        \log \left(
            c \sigma_{PN}(\bm{\rho}^m,\bar{\tilde{\bm{x}}},\bar{\tilde{\bm{s}}}^m) \lambda_Y^* \right) \leq 0,
        \end{split}
    \end{equation}
    where $\lambda_Y^*$ is the predefined target YLF or safety factor against yielding. The logarithm is employed to normalize the constraint. The constraint can be formulated such that the YLF is related to the global Buckling Load Factor (BLF), which i defined as $\lambda_B(\gamma_i(\bm{\rho}^m)) = 1/J^{KS}(\gamma_i(\bm{\rho}^m))$. Here $J^{KS}(\gamma_i(\bm{\rho}^m))$ is the KS aggregated eigenvalues scaling the stresses according to the critical BLF to ensure yield strength up to the buckling limit. The definition of $J^{KS}(\gamma_i(\bm{\rho}^m))$ can be found in \cite{Christensen2023}. This modification yields a constraint defined as
    \begin{equation}\label{eq:yieldConstraint}
        \begin{split}
        g_y^{const}(\bm{\rho}^m,\bar{\tilde{\bm{x}}},\bar{\tilde{\bm{s}}}^m,& \gamma_i(\bm{\rho}^m)) = \\ &\log \left(
        \dfrac{c \sigma_{PN}(\bm{\rho}^m,\bar{\tilde{\bm{x}}},\bar{\tilde{\bm{s}}}^m)}{J^{KS}(\gamma_i(\bm{\rho}^m))}  \right) \leq 0.
        \end{split}
    \end{equation}
    %
    
    \begin{Figure}
        \centering
            \begin{tikzpicture}
    \begin{axis}[
        xlabel = {\(\sigma_1/E_0\)},
        ylabel = {\(\sigma_2/E_0\)},
        legend pos = north east,
        xmin=-0.035, xmax=0.035,
        ymin=-0.035, ymax=0.035,
        width=6cm,   
        height=6cm,    
        xmajorgrids=true,
        ymajorgrids=true,
        legend style={at={(1.05,0.5)}, anchor=west}, 
        yticklabel style={
            /pgf/number format/fixed,
            /pgf/number format/precision=2,
        },
        scaled y ticks=false, 
        xticklabel style={
            /pgf/number format/fixed,
            /pgf/number format/precision=2,
        },
        scaled x ticks=false, 
    ]

    \draw[thick, black] (axis cs:-0.045,0) -- (axis cs:0.045,0); 
    \draw[thick, black] (axis cs:0,-0.045) -- (axis cs:0,0.045); 

    \addplot[
        color=mycolor1,
        dashed,
        line width = 2pt,
        draw opacity=1,
     ]
    table [x index=0, y index=1, col sep=comma] {figures/yieldSurface/fitContour_vMises_rho_4_theta_0.csv};
    \addlegendentry{$\rho_e = 0.4$}

    \addplot[
        color=mycolor2,
        dashed,
        line width = 2pt,
        draw opacity=1,
     ]
    table [x index=0, y index=1, col sep=comma] {figures/yieldSurface/fitContour_vMises_rho_5_theta_0.csv};
    \addlegendentry{$\rho_e = 0.5$}

    \addplot[
        color=mycolor3,
        dashed,
        line width = 2pt,
        draw opacity=1,
     ]
    table [x index=0, y index=1, col sep=comma] {figures/yieldSurface/fitContour_vMises_rho_6_theta_0.csv};
    \addlegendentry{$\rho_e = 0.6$}

    \addplot[
        color=mycolor1,
        line width = 2pt,
        draw opacity=0.4,
     ]
    table [x index=0, y index=1, col sep=comma] {figures/yieldSurface/fitContour_Willam-Warnke_rho_4_theta_0.csv};

    \addplot[
        color=mycolor2,
        line width = 2pt,
        draw opacity=0.4,
     ]
    table [x index=0, y index=1, col sep=comma] {figures/yieldSurface/fitContour_Willam-Warnke_rho_5_theta_0.csv};

    \addplot[
        color=mycolor3,
        line width = 2pt,
        draw opacity=0.4,
     ]
    table [x index=0, y index=1, col sep=comma] {figures/yieldSurface/fitContour_Willam-Warnke_rho_6_theta_0.csv};

    \end{axis}
\end{tikzpicture}%
        \captionof{figure}{Illustration of the yield failure surface (dashed lines) at $\rho_e= \{0.4, 0.5, 0.6\}$. For comparison, the full lines are the buckling failure surfaces.}
        \label{fig:yieldFailureSurface}
    \end{Figure}
    %
    The shape of the yield failure surface at three densities is illustrated in \cref{fig:yieldFailureSurface}, where it is compared to the buckling failure surfaces. The example in the figure uses a base material with yield limit $\sigma_0 = 0.1$. Biaxial compression and shear will be dominated by buckling failure for $\rho_e = 0.4$ for such a material. Increasing the density leads to more yield-dominated failure; for $\rho_e = 0.6$, yielding will be the dominant failure reason for all stress states.

    \subsection{De-homogenization}
    \label{sec:method:dehom}

    \begin{figure*}[t!]
        \centering
        \graphicspath{{figures/deHomogenization/}}
        \makebox[\textwidth][c]{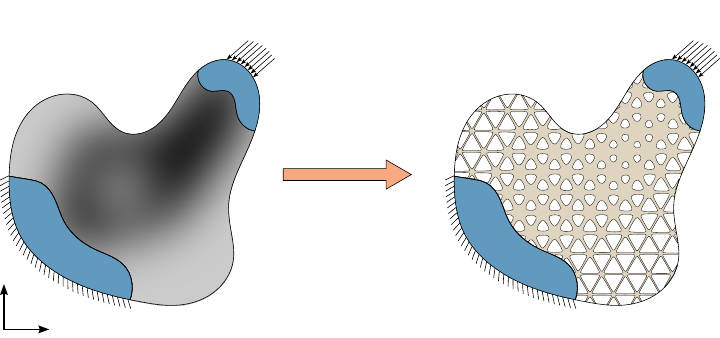}%
        \caption{Illustration a homogenized design field defined in the macroscale domain and corresponding de-homogenized structure varying locally according to the homogenized densities.}
        \label{fig:twoTermDeHom}
    \end{figure*}
    The de-homogenization using the two-parameter microstructure is straightforward. Given the globally defined periodic functions, \cref{eq:Fgeom}, \cref{eq:Fsharp} and \cref{eq:Ftriangle}, the repeating cells are automatically arranged. The two parameters defining the local microstructure are extracted using \cref{eq:etaBarFit} and \cref{eq:alphaFit} with the densities from the homogenized design field. The size of the de-homogenized microstructures is directly controlled through $d$.
    
    The de-homogenization is illustrated in \cref{fig:twoTermDeHom}, where an arbitrary homogenized design field is defined in the macroscale domain. The de-homogenized structure varies locally according to the homogenized design. The low-density regions have slender bars with small rounding at the corners. The denser regions consist of almost round holes and are more plate-like in their appearance. The coating with locally varying thickness from \cite{Christensen2023} can also be applied to this microstructure to achieve improved performance for increasing sizes of the microstructure.
        
    As mentioned in \cref{sec:method:microGeometry}, the de-homogenization can be accomplished either by thresholding (\cref{eq:dehom:heaviside}) or by extracting the contour (\cref{eq:dehom:contour}). In this work, both methods are performed in Matlab. The first method directly generates a mesh with discrete densities that can be evaluated using the same framework as used for the optimization. The second method generates the boundary of the de-homogenized design, which can be imported into commercial software such as COMSOL for further analysis. The advantage of this is that body-fitted meshes can be used, thus enabling advanced meshing techniques resulting in higher resolution at critical areas. 
    
    \section{Numerical results}
    \label{sec:results}
    
    This section demonstrates the suggested method by performing topology optimization of the classical L-beam design domain \cite{duysinx1998a}. This is done by first optimizing with the yield constraint formulation from \cref{eq:yieldConstraint}. Second, optimizations where the global BLF, Local Buckling Load Factor (LBLF) and YLF are maximized are performed using actual materials.
    
    All optimizations are performed on the L-beam design domain, illustrated in \cref{fig:LBeamYieldDomain}. To avoid "sticky" domain boundaries due to filtering of the design fields, the domain uses Robin type boundary conditions for the filters on the red colored boundaries. The remaining boundaries, use Neumann boundary conditions in the filter operations. The domain is fixed at the top and subjected to a distributed load on the upper part of the right side of the domain. The magnitude of the load is $f = -1 \times 10^{-3}$. A passive solid region is placed at the load with a thickness equivalent to the void indicator filter radius. The robust formulation uses $\Delta \eta = 0.01$, and the minimum allowed density is $x_{min} = 0.15$.
      %
    \begin{Figure}
        \centering
        \graphicspath{{figures/Domain/}}
\begingroup%
  \makeatletter%
  \providecommand\color[2][]{%
    \errmessage{(Inkscape) Color is used for the text in Inkscape, but the package 'color.sty' is not loaded}%
    \renewcommand\color[2][]{}%
  }%
  \providecommand\transparent[1]{%
    \errmessage{(Inkscape) Transparency is used (non-zero) for the text in Inkscape, but the package 'transparent.sty' is not loaded}%
    \renewcommand\transparent[1]{}%
  }%
  \providecommand\rotatebox[2]{#2}%
  \newcommand*\fsize{\dimexpr\f@size pt\relax}%
  \newcommand*\lineheight[1]{\fontsize{\fsize}{#1\fsize}\selectfont}%
  \ifx\svgwidth\undefined%
    \setlength{\unitlength}{144.10072387bp}%
    \ifx\svgscale\undefined%
      \relax%
    \else%
      \setlength{\unitlength}{\unitlength * \real{\svgscale}}%
    \fi%
  \else%
    \setlength{\unitlength}{\svgwidth}%
  \fi%
  \global\let\svgwidth\undefined%
  \global\let\svgscale\undefined%
  \makeatother%
  \begin{picture}(1,1.00332877)%
    \lineheight{1}%
    \setlength\tabcolsep{0pt}%
    \put(0.84538093,0.53963094){\color[rgb]{0,0,0}\makebox(0,0)[lt]{\lineheight{1.25}\smash{\begin{tabular}[t]{l}$f$\end{tabular}}}}%
    \put(0,0){\includegraphics[width=\unitlength,page=1]{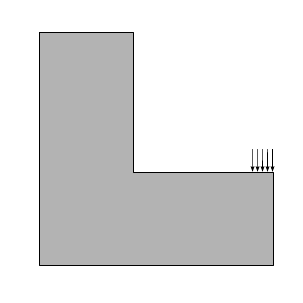}}%
    \put(0.4965933,0.00000007){\makebox(0,0)[lt]{\lineheight{1.25}\smash{\begin{tabular}[t]{l}1\end{tabular}}}}%
    \put(0,0){\includegraphics[width=\unitlength,page=2]{LDomainYield.pdf}}%
    \put(0.00867123,0.47582144){\makebox(0,0)[lt]{\lineheight{1.25}\smash{\begin{tabular}[t]{l}1\end{tabular}}}}%
    \put(0,0){\includegraphics[width=\unitlength,page=3]{LDomainYield.pdf}}%
    \put(0.96922931,0.25571268){\makebox(0,0)[lt]{\lineheight{1.25}\smash{\begin{tabular}[t]{l}0.4\end{tabular}}}}%
    \put(0.23636323,0.97242581){\makebox(0,0)[lt]{\lineheight{1.25}\smash{\begin{tabular}[t]{l}0.4\end{tabular}}}}%
    \put(0.71210333,0.57250895){\color[rgb]{0,0,0}\makebox(0,0)[lt]{\lineheight{1.25}\smash{\begin{tabular}[t]{l}$\Omega_P$\end{tabular}}}}%
    \put(0.70418991,0.3649889){\color[rgb]{0,0,0}\makebox(0,0)[lt]{\lineheight{1.25}\smash{\begin{tabular}[t]{l}$r_s$\end{tabular}}}}%
    \put(0.74725956,0.22732596){\color[rgb]{0,0,0}\makebox(0,0)[lt]{\lineheight{1.25}\smash{\begin{tabular}[t]{l}0.06\end{tabular}}}}%
    \put(0,0){\includegraphics[width=\unitlength,page=4]{LDomainYield.pdf}}%
    \put(0.27160125,0.29368951){\color[rgb]{0,0,0}\makebox(0,0)[lt]{\lineheight{1.25}\smash{\begin{tabular}[t]{l}$\Omega_A$\end{tabular}}}}%
    \put(0,0){\includegraphics[width=\unitlength,page=5]{LDomainYield.pdf}}%
  \end{picture}%
\endgroup%

        \captionof{figure}{Illustration of the L-beam design domain. The domain is fixed at the top and subjected to a load at the top right part. Robin type boundary conditions are used to avoid "sticky" boundaries.}
        \label{fig:LBeamYieldDomain}
    \end{Figure}
        
    \subsection{Yield Stress Constraint}\label{sec:results:yieldStressConstraint}
    This section presents numerical results using the yield stress constraint formulation defined by \cref{eq:yieldConstraint}. The active domain is discretized using $100 \times 100$ bilinear quadrilateral elements. The void indicator filter radius $r_s = 5$ elements. The density filter radius is $r_x=1.5$ elements. The material is modeled using $E_0 = 1$, $\nu = 1/3$, and $\sigma_0 = 0.366$. The optimization problem is formulated as
    \begin{alignat}{3}\label{eq:optProblem:LBeamYieldDomain:1}
        \min\limits_{\mathbf{x},\mathbf{s}} \ &: \  g_c(\boldsymbol{\rho}^e) + 
        \Gamma \frac{V^e_{\Omega,struct}(\bm{s})}{V_{\Omega}}, &\quad & \\ 
        \textrm{s.t.} &: \ g_{\lambda}(\bm{\rho}^m) 
        \leq 0, &\ & m\in\{e,i,d\}, \ \gamma_i\in\mathcal{B},\label{eq:optProblem:LBeamYieldDomain:2}\\ 
         &: \ g_V(\boldsymbol{\rho}^d) \leq 0, &\ & \label{eq:optProblem:LBeamYieldDomain:3}\\ 
        &: \ g_y(\bm{\rho}^m,\bar{\tilde{\bm{x}}},\bar{\tilde{\bm{s}}}^m,\gamma_i(\bm{\rho}^m)) \leq 0, &\ &m\in\{e,i,d\}, \ \gamma_i\in\mathcal{B}, \label{eq:optProblem:LBeamYieldDomain:4}\\
        &: \ g_s(\bm{\rho}^m,\bar{\tilde{\bm{x}}},\bar{\tilde{\bm{s}}}^m,\gamma_i(\bm{\rho}^m)) \leq 0, &\ &m\in\{e,i,d\}, \ \gamma_i\in\mathcal{B}, \label{eq:optProblem:LBeamYieldDomain:5}\\
        &: \ \rho_e^m = \bar{\tilde{x}}_e \bar{\tilde{s}}_e^m, &\ &m\in\{e,i,d\}, \ \forall e, \label{eq:optProblem:LBeamYieldDomain:6}\\
        &: \ x_{min} \leq x_e \leq 1, &\ &\forall e, \label{eq:optProblem:LBeamYieldDomain:7}\\
        &: \ 0 \leq s_e \leq 1, &\ &\forall e. \label{eq:optProblem:LBeamYieldDomain:8}
    \end{alignat}
    Here \cref{eq:optProblem:LBeamYieldDomain:1} is the objective which aims at maximizing the stiffness of the structure.  The objective is augmented with a term to penalize the volume of the eroded indicator field $V^e_{\Omega,struct}(\bm{s})$ with a factor $\Gamma$ to avoid a "plateau" of densities equal to $x_{min}$ along outer boundaries when optimizing for minimum compliance \cite{Giele2021}. The buckling constraint in \cref{eq:optProblem:LBeamYieldDomain:2} is defined to satisfy a pre-defined value $\lambda^*$ on all of the robust projections. The definition of \cref{eq:optProblem:LBeamYieldDomain:2} is defined in \cite{Christensen2023}. The volume constraint in \cref{eq:optProblem:LBeamYieldDomain:3} is defined following \cite{Lazarov2016}. The multiscale yield constraint is enforced in \cref{eq:optProblem:LBeamYieldDomain:4} following the definition in \cref{eq:yieldConstraint}. The buckling stress constraint in \cref{eq:optProblem:LBeamYieldDomain:5} is defined in \cite{Christensen2023} with the modification in \cref{eq:AndersenBucklingInterpolationModified}. Finally, the physical densities $\rho_e^m$ come from $\bar{\tilde{x}}_e$ and $\bar{\tilde{s}}_e^m$. The box limits of the design fields are defined in \cref{eq:optProblem:LBeamYieldDomain:7} and \cref{eq:optProblem:LBeamYieldDomain:8}.
    
    
    The optimization problem is solved using the MMA optimizer \cite{Svanberg1987} with a continuation strategy for the sharpness parameter $\beta$. For this complex optimization problem, which is subject to many constraints, the continuation can be rather sensitive and prone to local minima. Therefore, the continuation strategy here is performed slowly and $\beta$ is updated after the first 125 iterations and at every 75 iteration after that. It is updated as $\beta^{n+1} = 1.3 \beta^{n}$, until the maximum value of 256. The continuation strategy is used to ensure that the optimization problem converges to a feasible solution. Alternatively, the authors have experienced fairly good convergence with the automatic projection method suggested by Dunning \cite{Dunning2024} and Dunning and Wein \cite{Dunning2025}, even with the number of constraints present in \cref{eq:optProblem:LBeamYieldDomain:1}--\cref{eq:optProblem:LBeamYieldDomain:8}.
    
    Four variations of the optimization problem are tested. All of the variations are allowed to use a volume fraction of $V^*_i = 0.35$ in the intermediate design. The four tests are compliance minimization with a volume constraint (CV), BLF maximization with compliance and a volume constraint (BCV), BLF maximization with compliance, volume, and a local buckling constraint (BCVS), and the last variant (CBVSY) uses the optimization problem as stated in \cref{eq:optProblem:LBeamYieldDomain:1}--\cref{eq:optProblem:LBeamYieldDomain:8}.
    
    The compliance from the CV design is used as the reference, and the compliance constraint in the BVC and BVCS optimizations is defined as $C^*_ e = 1.25 C_{CV}$ i.e., the compliance is allowed to increase by \SI{25}{\percent}. The buckling constraint on the CBVSY problem is based on the BLF of the BCVS design, i.e., $\lambda^* = \lambda_{1,BCVS}$.

    \subsubsection{Optimized Designs}
\begin{figure*}
        \centering
        \graphicspath{{figures/DesignsHom/}}
        \makebox[\textwidth][c]{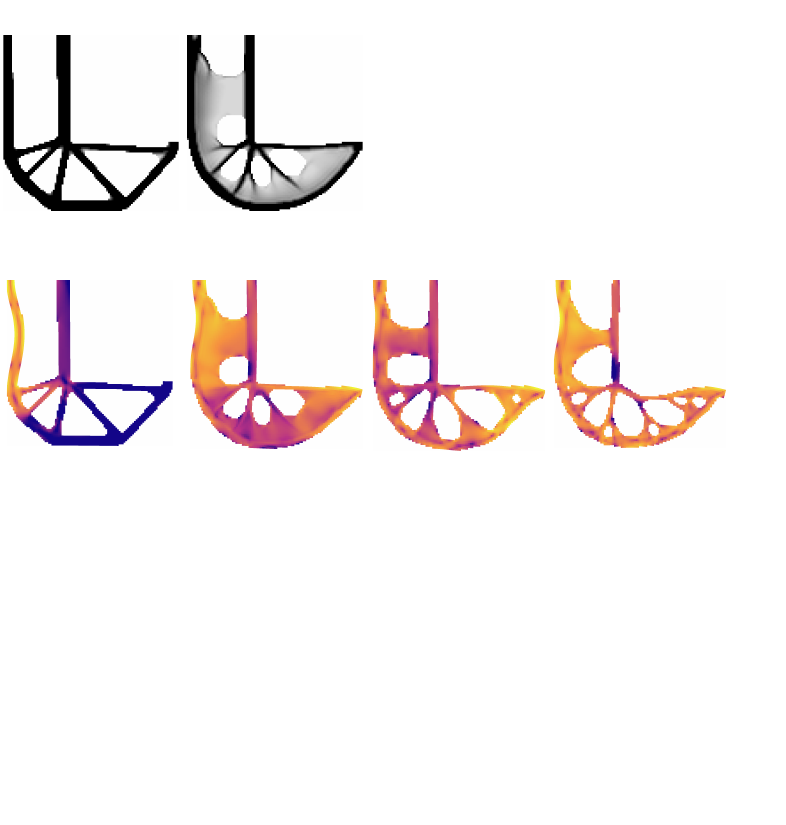
        \phantomsubcaption
        \label{fig:yieldHomogenizedModes:1}
        \phantomsubcaption
        \label{fig:yieldHomogenizedModes:2}
        \phantomsubcaption
        \label{fig:yieldHomogenizedModes:3}
        \phantomsubcaption
        \label{fig:yieldHomogenizedModes:4}
        }%
        \caption{The four homogenized designs and their respective buckling modes and von Mises stress fields. The colormap in the center row indicates the normalized strain energy density $\log_{10}(W_e/W_{max})$. The colormap in the bottom row is the microscale von Mises stress field at the critical load. (a) CV, (b) BCV, (c) BCVS, (d) CBVSY.}
        \label{fig:yieldHomogenizedModes}
    \end{figure*}

    The four optimized designs are shown in \cref{fig:yieldHomogenizedModes} with the compliance of the eroded designs normalized with respect to that of the CV design. The figure also includes the critical buckling mode with the lowest BLF of the three robust realizations as well as the von Mises stress distributions at the critical load. To account for the stress concentrations on the microscale level, the stresses are scaled according to stress concentration factors from \cref{eq:relaxedYieldLimit}
    \begin{equation}
        \sigma_{VM,m}(\bar{\tilde{x}}_e) = \frac{1}{{\sigma}_{lim}(\bar{\tilde{x}}_e)} \sigma_{VM,M}(\bar{\tilde{x}}_e),
    \end{equation}
    where subscripts $m$ and $M$ indicate microscale and macroscale, respectively. Furthermore, $\sigma_0 = 1$ to use \cref{eq:relaxedYieldLimit} only as a scaling of the stresses. The macroscale von Mises stress $\sigma_{VM,M}$ is scaled with $\lambda_1$ to get the stress at the critical load. 
    
    Taking a look at the optimized designs shows that the compliance-minimized design is a single-scale structure even when allowed to use multiscale. This result aligns perfectly with the conclusions from \cite{Christensen2023}. Small regions of intermediate densities around the corners of the macroscale structure exist. These regions are a result of the two filter radii used in the multifield formulation. The macroscale design is governed by the indicator filter radius $r_s$, and the infill densities are governed by $r_x$. The optimizer exploits this to use low-density material in regions where sharp corners are preferred but restricted by $r_s$ in the macroscale topology.
    
    The long, slender bars in the CV design are prone to buckling at low loads. The buckling mode shows that buckling occurs in the leftmost bar in the design, which is in compression and has the highest length-to-thickness ratio. 
    For the CV design, the maximum von Mises stress is below the yield stress of the base material ($\sigma_0 = 0.366$), even with the sharp re-entrant corner in the design. This is because the low stability of the design results in buckling at low stresses hence stresses do not exeed the yield limit because  the problem is buckling dominated.
    
    The BCV design in \cref{fig:yieldHomogenizedModes:2} utilizes multiscale material to stabilize the long solid sections. The design resembles a coated infill structure, resulting in a significantly higher BLF compared to the CV design. However, the high BLF combined with a sharp re-entrant corner leads to a high stress concentration. The maximum microscale von Mises stress is $\sigma_{VM} = 1.73$, which is significantly above the yield strength of the material. This stress is found in the non-solid element at the re-entrant corner, introducing a microscale stress concentration. This intermediate density element is present in all designs where the yield stress constraint is not included. On top of this, the low-density infill is likely to experience local buckling with a risk of reducing the load-carrying capabilities \cite{Christensen2023}.
    
    The BVCS design prevents local buckling, and the effect on the design is seen in \cref{fig:yieldHomogenizedModes:3}. More macroscale holes are added, and the infill density is increased. The BLF is more than six times higher than for the CV design, but the design still features a sharp, re-entrant corner. The maximum von Mises stress in the structure is $\sigma_{VM} = 1.48$, which significantly exceeds the yield strength of the base material. As for the BCV design the stress concentration is amplified by the element with the intermediate density at the re-entrant corner.
    
    The final design is the CBVSY design shown in \cref{fig:yieldHomogenizedModes:4}. The BLF is constrained to be at least the same level as the BLF of the BCVS design. This results in increased compliance to fulfill all the remaining constraints. Again, intermediate material is used to support solid load-carrying regions. The re-entrant corner is now rounded to reduce the stress concentrations. The maximum von Mises stress is $\sigma_{VM} = 0.37 = 1.006 \sigma_0$. Thus, the design successfully meets the optimization criteria by reducing the maximum von Mises stress by \SI{75}{\percent} compared to the BCVS design.
    
    \subsubsection{De-homogenized Designs}
\begin{figure*}
        \centering
        \graphicspath{{figures/Dehom/}}
        \makebox[\textwidth][c]{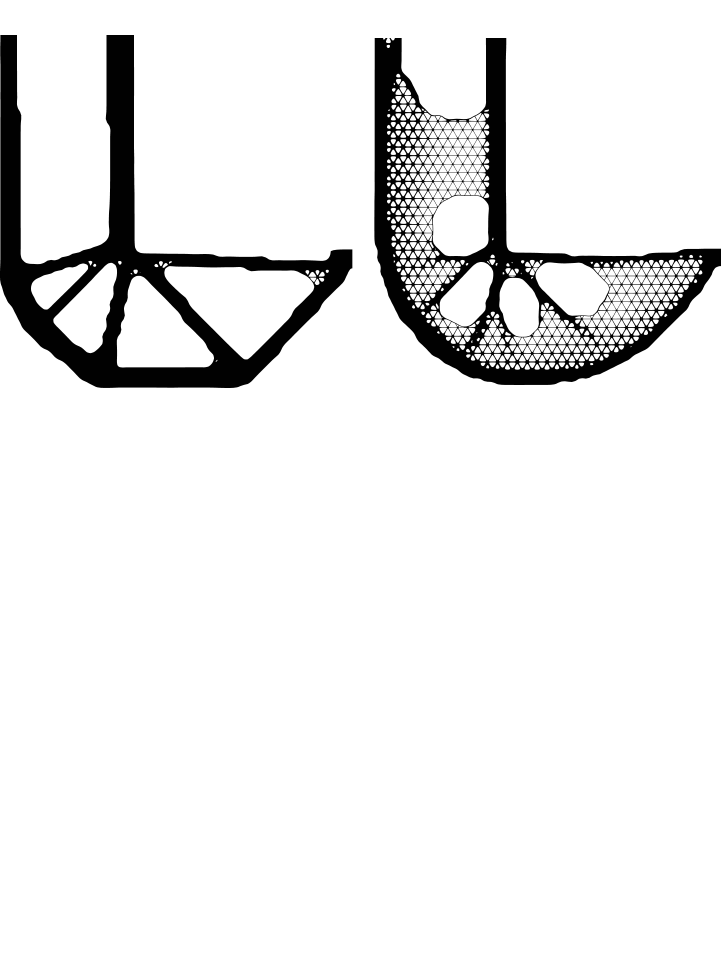
        \phantomsubcaption
        \label{fig:yieldDehom:1}
        \phantomsubcaption
        \label{fig:yieldDehom:2}
        \phantomsubcaption
        \label{fig:yieldDehom:3}
        \phantomsubcaption
        \label{fig:yieldDehom:4}
        }%
        \caption{The four designs de-homogenized using the two-parameter microstructure. (a) CV, (b) BCV, (c) BCVS, (d) CBVSY.}
        \label{fig:yieldDehom}
    \end{figure*}

    The designs from \cref{fig:yieldHomogenizedModes} are de-homogenized using the two-parameter parameterization introduced in \cref{sec:method:microGeometry}. The size of the microstructure is $d=0.03$, and the boundary of the structure is extracted using \cref{eq:dehom:contour}. The de-homogenized structures are imported in COMSOL 6.1 and bodyfitted meshes are used for post-evaluation. The structures are presented in \cref{fig:yieldDehom}.
    
    Again, the de-homogenized CV design is predominantly single-scale, except for the small porous regions at the corners. The size of the microstructure combined with the coating results in a slight smoothing of the re-entrant corner, which will reduce stress concentrations.
    
    The de-homogenized BCV structure in \cref{fig:yieldDehom:2} exhibits the same rounding at the re-entrant corner as the CV design. The figure nicely illustrates how the two-parameter microstructure adapts hole size and shape according to the local infill densities.
    
    The BCVS design is de-homogenized in \cref{fig:yieldDehom:3}. The structure has dense microstructure in regions subject to high compression. The microstructure is less dense in the more shear-dominated regions, and the bars in the microstructure are thinner.
    
    Finally, the CBVSY design, optimized with a constraint on the yield stress, is de-homogenized in \cref{fig:yieldDehom:4}. As with the BCV and BCVS designs, the microstructure varies locally according to the local densities. Most importantly, the large, smooth rounding with no porous material at the re-entrant corner is crucial for reducing stress concentrations.
    
    \subsubsection{Numerical Post-evaluation}
    The de-homogenized structures from \cref{fig:yieldDehom} are numerically evaluated using COMSOL 6.1. A body-fitted triangular mesh is used with an element size between $2\times 10^{-4}$\ --\ $5\times 10^{-3}$. The compliance, BLF, and yield strength are evaluated and compared to the homogenized designs. All the data is presented in \cref{tab:yieldDehomData}, where it is compared to the homogenized designs. All data in the table is for the intermediate design, which is the one that is de-homogenized. The maximum von Mises stress in the de-homogenized structures is evaluated at the critical load of both the homogenized and de-homogenized structures.
    %
    \begin{center}
        \captionof{table}{Homogenized and de-homogenized data for the four optimized intermediate L-beam designs. $\lambda_1^H$ and $\lambda_1^D$ are the critical BLFs of the homogenized and de-homogenized designs.}\label{tab:yieldDehomData}
        \begin{tabular}{lrrrr}
           & \textbf{CV} & \textbf{BCV} & \textbf{BCVS} & \textbf{CBVSY} \\
        \midrule
        & \multicolumn{4}{c}{Homogenized Analysis} \\
        \midrule
        $V$ & 0.35 & 0.35 & 0.35 & 0.35 \\
        $C$ ($\times10^{-4}$) & 1.59 & 1.98 & 1.98 &  2.15 \\
        $\lambda_1^H$ & 0.73 & 5.28 & 4.6 & 4.62 \\
        $\max{(\sigma_{VM,m})}$ & 0.18 & 1.73 & 1.48 & 0.37 \\
        \midrule
        & \multicolumn{4}{c}{De-homogenized Analysis} \\
        \midrule
        $V$ & 0.345 & 0.352 & 0.355 & 0.356 \\
        $C$ ($\times10^{-4}$)& 1.63 & 2.00 & 1.96 & 2.12 \\
        $\lambda_1^D$ & 0.63 & 0.28 & 4.53 & 4.56 \\
        $\max{(\sigma_{VM} \lambda_1^H)}$ & 0.09 & 0.81 & 0.70 & 0.43 \\
        $\max{(\sigma_{VM} \lambda_1^D)}$ & 0.08 & 0.04 & 0.69 & 0.42 \\
        \bottomrule
        \end{tabular}
    \end{center}
    
    The data in the table shows good correlation between the homogenized and de-homogenized designs for both volume and compliance, and is further discussed in the following.
    
\begin{figure*}
        \centering
        \graphicspath{{figures/DeHomBuckling/}}
        \makebox[\textwidth][c]{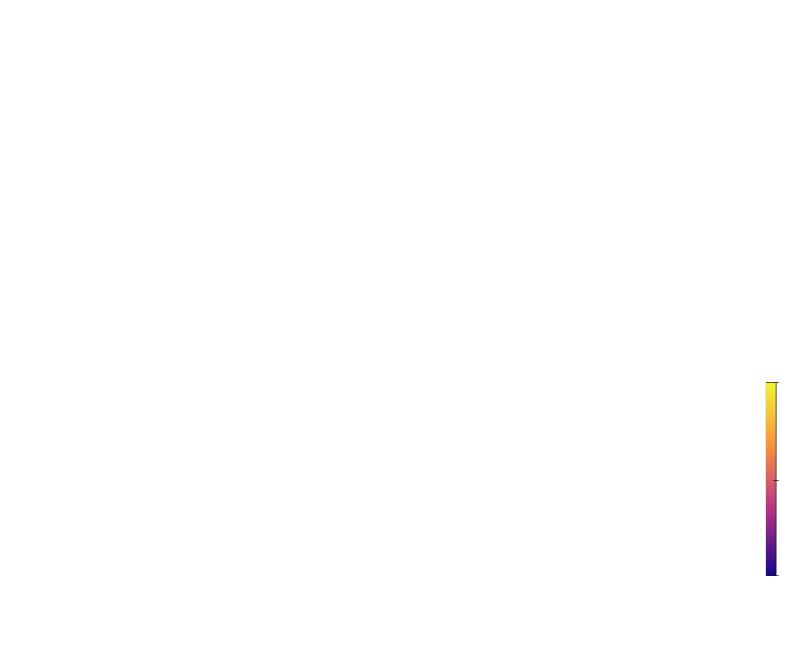
        \phantomsubcaption
        \label{fig:dehomBuckling:1}
        \phantomsubcaption
        \label{fig:dehomBuckling:2}
        \phantomsubcaption
        \label{fig:dehomBuckling:3}
        \phantomsubcaption
        \label{fig:dehomBuckling:4}
        \phantomsubcaption
        \label{fig:dehomBuckling:5}
        }%
        \caption{The critical buckling modes for the four de-homogenized designs. The colormap shows the normalized strain energy density $\log_{10}(W_e/W_{max})$. (a) CV, (b) BCV, (c) BCVS, (d) First local buckling modes of the CBVSY design, (e) Third global mode of the CBVSY design.}
        \label{fig:dehomBuckling}
    \end{figure*}

    \paragraph{Critical Buckling Load}
    The critical buckling modes for the four de-homogenized structures are presented in \cref{fig:dehomBuckling}. The BLFs and buckling modes of the CV and BCVS designs are clearly global. 
    The BLF of the BCV design is significantly lower than the one estimated by the homogenized analysis. As shown in \cref{fig:dehomBuckling:2}, this discrepancy is due to local buckling within the microstructure. The homogenized optimization exploits the low-density infill to enhance buckling stability. However, the resulting slender microstructures have very little resistance to buckling and fail at much lower loads than anticipated. This also explains the difference in \cref{tab:yieldDehomData}.
    
    The first buckling mode of the CBVSY design is local and is shown in \cref{fig:dehomBuckling:4} with clear buckling of the microstructure. However, the BLF is only \SI{1}{\percent} lower than that estimated by the homogenized analysis. The third mode of the CBVSY design is a global mode, shown in \cref{fig:dehomBuckling:5} and has a BLF that is only \SI{5}{\percent} higher than the first mode. This means that the first modes are almost active at the same time, which aligns perfectly with the formulation of the multiscale buckling constraint. If global failure was desired, the local buckling constraint could be updated with a safety factor similar to the approach suggested in  \cite{Christensen2025}.

    \paragraph{Yield Stress}
    
\begin{figure*}
        \centering
        \graphicspath{{figures/DeHomYield/}}
        \makebox[\textwidth][c]{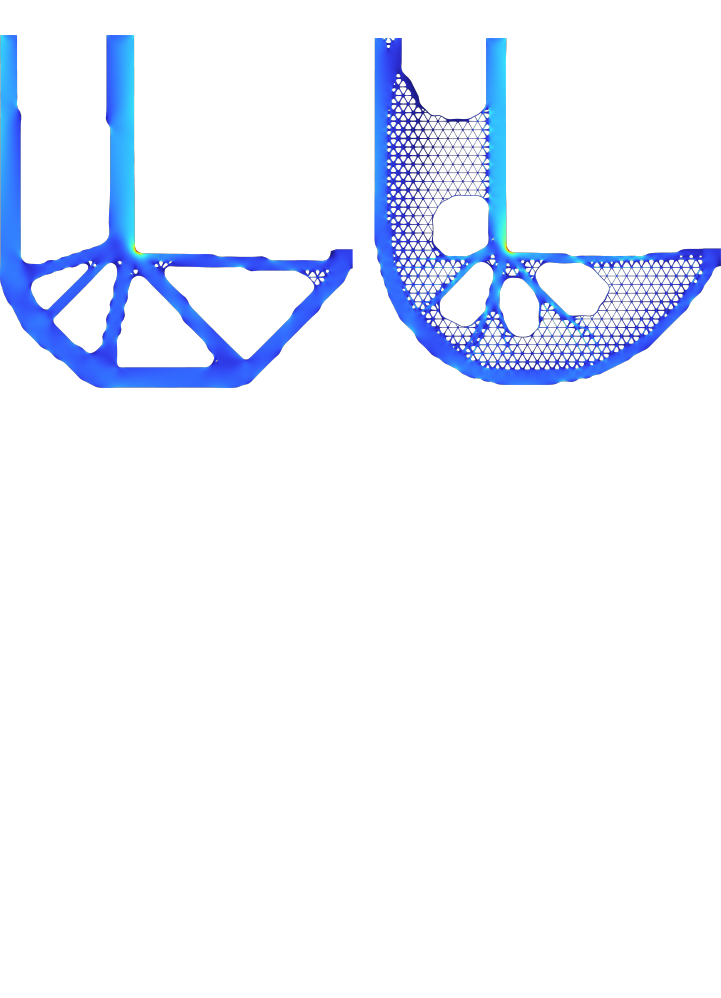
        \phantomsubcaption
        \label{fig:dehomYield:1}
        \phantomsubcaption
        \label{fig:dehomYield:2}
        \phantomsubcaption
        \label{fig:dehomYield:3}
        \phantomsubcaption
        \label{fig:dehomYield:4}
        }%
        \caption{The von Mises stress distributions for the four de-homogenized designs. The colormap shows the von Mises stresses scaled with the BLF from the homogenized design $\lambda_1^H$. (a) CV, (b) BCV, (c) BCVS, (d) CBVSY.}
        \label{fig:dehomYield}
    \end{figure*}

    The von Mises stress fields in the four de-homogenized structures are shown in \cref{fig:dehomYield}. The stresses are scaled with the de-homogenized designs' critical BLFs $\lambda_1^D$. The colorbars indicate the maximum von Mises stress in the structures. The maximum von Mises stress scaled with critical BLF $\lambda_1^H$ of the de-homogenized design is shown in \cref{tab:yieldDehomData}.
    
    The CV structure in \cref{fig:dehomYield:1} has a stress concentration at the sharp, re-entrant corner. The zoom-in on the corner shows high stresses along the rounded outer boundary. The maximum von Mises stress is $\sigma_{VM} = 0.08$, approximately half of the estimated worst-case stress in the homogenized analysis. The reason for the difference lies in the de-homogenization at the re-entrant corner. \cref{fig:yieldHomogenizedModes:1} shows a slightly porous material at the re-entrant corner. This is not present in the de-homogenized design, which results in a more rounded corner and less stress concentration than if holes were present in the structure. The reason for not seeing holes in the de-homogenized structures comes down to the chosen size of the microstructure and the coating around all void interfaces. The microstructure is too large to capture the holes, and the coating is too thick to allow for holes to be present at the re-entrant corner in the de-homogenized design.
    
\begin{figure*}
        \centering
        \graphicspath{{figures/DeHomYield/}}
        \makebox[\textwidth][c]{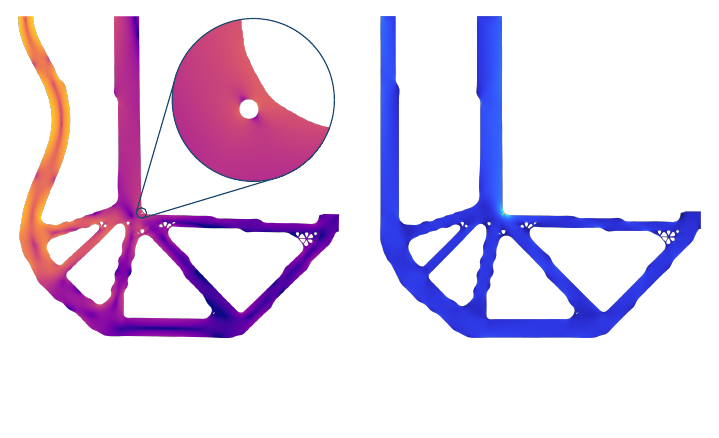
        \phantomsubcaption
        \label{fig:dehomYieldHole:1}
        \phantomsubcaption
        \label{fig:dehomYieldHole:2}
        }%
        \caption{Numerical results of the CV design with a hole at the re-entrant corner. (a) Critical buckling mode with $\lambda_1 = 0.63$, (b) The von Mises stress distribution.}
        \label{fig:dehomYieldHole}
    \end{figure*}

    To examine the influence of having porous material at the re-entrant hole, an additional analysis is made. \cref{fig:dehomYieldHole} shows the CV design with a hole at the re-entrant corner. The critical buckling mode is shown in \cref{fig:dehomYieldHole:1} with $\lambda_1 = 0.63$, i.e., the hole has no effect on the stability of the structure. The von Mises stress distribution is shown in \cref{fig:dehomYieldHole:2}. The stress field is similar to the CV design, but the stress concentration at the re-entrant corner is significantly higher and located on the boundary of the hole instead of the outer boundary at the corner. The maximum von Mises stress is $\sigma_{VM} = 0.12$, which is closer to the worst-case estimate. This shows that the porous material at the re-entrant corner introduces a risk of even higher stress concentrations than fully solid material.
    
    The maximum von Mises stress in both cases is still below the base material's yield stress, regardless of whether the hole is present or not. This is due to the low stability of the design, which results in buckling at low loads. Therefore, the stress concentrations are not high enough to exceed the yield limit of the base material.
    
    The stress field of the BCV design is shown in \cref{fig:dehomYield:2}. The de-homogenized structure features the same sharp rounding at the re-entrant corner as the CV design. With the very low BLF of the local mode in the de-homogenized design $\lambda_1^D = 0.28$, the stresses at the critical buckling load are far less than the yield strength of the base material. However, if the stresses are scaled with the homogenized BLF the stresses at the critical buckling load exceed the yield strength of the base material significantly. Thus, the BCV structure highlights the need for not only a local buckling constraint, but also a yield stress constraint.
    
    The BCVS structure has a small rounding with solid material at the re-entrant corner, similar to the CV and BCV designs. With a maximum stress of $\sigma_{VM} = 0.70$ this means that the stresses at the critical load considerably exceed the base material's yield limit regardless of whether $\lambda_1^H = 4.6$ or $\lambda_1^D = 4.53$ are used to scale the stresses.
    
    The CBVSY structure, where the stresses are constrained, is shown in \cref{fig:dehomYield:4}. The less blue colormap indicates that stress concentrations are reduced compared to the three previous designs. The stresses are distributed much more smoothly around the re-entrant corner because of the large rounding. With a maximum von Mises stress of 0.43 when using the homogenized scaling and 0.42 for the de-homogenized scaling, the stresses are reduced by \SI{40}{\percent} compared to the BCVS design. 
    
    The maximum von Mises stress at the critical load, estimated by the homogenized analyses $\lambda_1^D$, is just \SI{14}{\percent} higher than the yield strength of the base material. Given the coarse mesh used for the homogenized analysis, where stresses are evaluated at element centers, this can be seen as a successful result, but also a topic for future work. A possible improvement for the future is to look at alternatives to evaluating stress at the element center or testing mesh refinement. 
    
\begin{figure*}
        \centering
        \graphicspath{{figures/LBeamFine/}}
        \makebox[\textwidth][c]{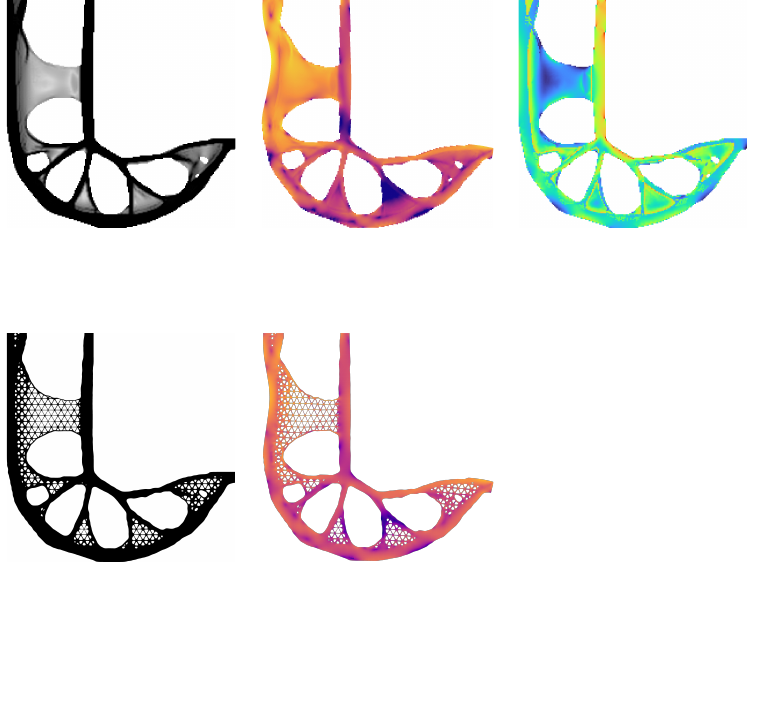
        \phantomsubcaption
        \label{fig:LBeamFine:1}
        \phantomsubcaption
        \label{fig:LBeamFine:2}
        \phantomsubcaption
        \label{fig:LBeamFine:3}
        }
        \caption{Optimization result performed on a fine mesh. (a) The optimised designs, (b) The critical buckling mode, (c) The von Mises stress field at the critical load.}
        \label{fig:LBeamFine}
    \end{figure*}

    As a quick test of this, the same optimization is performed with $200 \times 200$ elements. The optimized design is presented in \cref{fig:LBeamFine}. The homogenized and de-homogenized designs in \cref{fig:LBeamFine:1} are very similar to the design from the optimization on the coarse mesh. The same is the case for the BLF and buckling mode in \cref{fig:LBeamFine:2}. For this design the first buckling mode is global in the de-homogenized design. The von Mises stress field in \cref{fig:LBeamFine:3} shows that the stress concentrations are reduced compared to the design from the coarse mesh. The maximum von Mises stress in the de-homogenized design is $\sigma_{VM} = 0.35$ which is below the yield stress of the base material. This shows that the stress concentrations are better captured by refining the mesh.

\begin{figure*}
        \centering
        \graphicspath{{figures/DeHomYieldThreshold/}}
        \makebox[\textwidth][c]{
\begingroup%
  \makeatletter%
  \providecommand\color[2][]{%
    \errmessage{(Inkscape) Color is used for the text in Inkscape, but the package 'color.sty' is not loaded}%
    \renewcommand\color[2][]{}%
  }%
  \providecommand\transparent[1]{%
    \errmessage{(Inkscape) Transparency is used (non-zero) for the text in Inkscape, but the package 'transparent.sty' is not loaded}%
    \renewcommand\transparent[1]{}%
  }%
  \providecommand\rotatebox[2]{#2}%
  \newcommand*\fsize{\dimexpr\f@size pt\relax}%
  \newcommand*\lineheight[1]{\fontsize{\fsize}{#1\fsize}\selectfont}%
  \ifx\svgwidth\undefined%
    \setlength{\unitlength}{357.65387828bp}%
    \ifx\svgscale\undefined%
      \relax%
    \else%
      \setlength{\unitlength}{\unitlength * \real{\svgscale}}%
    \fi%
  \else%
    \setlength{\unitlength}{\svgwidth}%
  \fi%
  \global\let\svgwidth\undefined%
  \global\let\svgscale\undefined%
  \makeatother%
  \begin{picture}(1,1.38383173)%
    \lineheight{1}%
    \setlength\tabcolsep{0pt}%
    \put(0,0){\includegraphics[width=\unitlength,page=1]{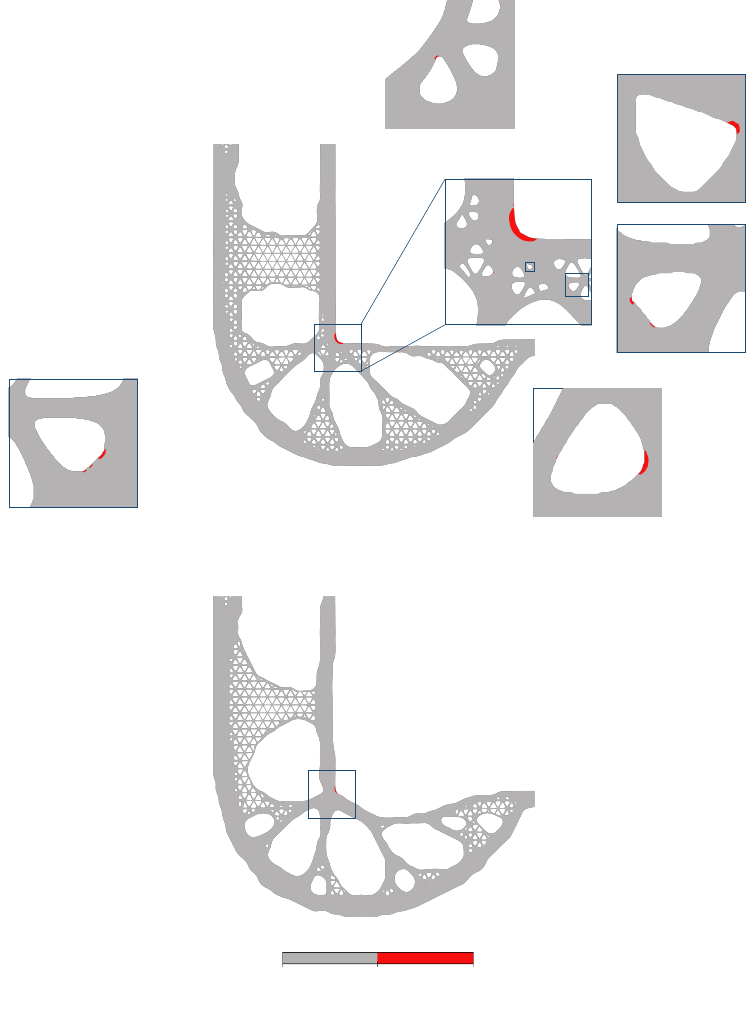}}%
    \put(0.30086817,0.05416457){\color[rgb]{0,0,0}\makebox(0,0)[lt]{\lineheight{1.25}\smash{\begin{tabular}[t]{l}$\sigma_{VM}<\sigma_0$\end{tabular}}}}%
    \put(0.56046437,0.05416457){\color[rgb]{0,0,0}\makebox(0,0)[lt]{\lineheight{1.25}\smash{\begin{tabular}[t]{l}$\sigma_{VM}>\sigma_0$\end{tabular}}}}%
    \put(0.48522931,0.05416457){\color[rgb]{0,0,0}\makebox(0,0)[lt]{\lineheight{1.25}\smash{\begin{tabular}[t]{l}$\sigma_0$\end{tabular}}}}%
    \put(0.48978316,0.65526354){\color[rgb]{0,0,0}\makebox(0,0)[lt]{\lineheight{1.25}\smash{\begin{tabular}[t]{l}(a)\end{tabular}}}}%
    \put(0.48960072,0.00218963){\color[rgb]{0,0,0}\makebox(0,0)[lt]{\lineheight{1.25}\smash{\begin{tabular}[t]{l}(b)\end{tabular}}}}%
    \put(0,0){\includegraphics[width=\unitlength,page=2]{dehomYieldThreshold.pdf}}%
  \end{picture}%
\endgroup%

        \phantomsubcaption
        \label{fig:dehomYieldThreshold:1}
        \phantomsubcaption
        \label{fig:dehomYieldThreshold:2}
        }
        \caption{Illistration of the violation of the yield stress constraint. The colormap shows the von Mises stresses with red indicating regions with $\sigma_{VM}\geq\sigma_0$ and gray indicating regions with $\sigma_{VM}<\sigma_0$. (a) The BCVS design, (b) The CBVSY design.}
        \label{fig:dehomYieldThreshold}
    \end{figure*}

    
    Taking a closer look at the CBVSY design, in \cref{fig:yieldHomogenizedModes:4} and \cref{fig:yieldDehom:4}, reveals that solid material is used where the stress concentration is highest. A natural conclusion to this could be that a traditinal macroscale stress constraint is sufficient to prevent material yielding. However, \cref{fig:dehomYieldThreshold:1} shows that the yield stress constraint is violated in the BCVS design both at the re-entrant corner, but also along the edges of the holes in the microstructure. Therefore, it is crucial that the yield stress constraint handles stresses on both micro and macroscale levels. The effect of this is confirmed in \cref{fig:dehomYieldThreshold:2}, where the yield stress constraint is not violated anywhere in the microstructure of the CBVSY design. The only violation of the yield limit is at the re-entrant corner. However, size of the violation is significantly reduced compared to the BCVS design and can be explained by the simple stress integration discussed above.
    
    In summary, the results of the four optimized designs demonstrate that homogenized optimization produces structures that are both stable and effective in reducing stress concentrations. Earlier work by Christensen et al. \cite{Christensen2023} highlighted improved BLF performance for de-homogenized designs. This study confirms that, even with the assumption of separation of scales, de-homogenization at relatively coarse length scales closely aligns with predictions from homogenized analyses. Deviations of at most \SI{15}{\percent} in stress predictions are attributed to the coarse mesh used in the homogenized analysis. When employing a finer mesh, the predicted stresses are accurate enough to satisfy the constraint even after de-homogenization.

    \subsection{Physical Material Dependency}\label{sec:results:materialTest}
    This section presents optimizations where BLF, LBLF and YLFs are all maximized. This time, we use physical material properties of five widely different materials to demonstrate how optimized designs dependent on material choice. The active domain is discretized using $200 \times 200$ bilinear quadrilateral elements. The void indicator filter radius $r_s = 10$ elements. The density filter radius is $r_x=1.5$ elements. The properties of the five materials are defined in \cref{tab:matProb}. In the optimization, the material properties are normalized with the Young's modulus, i.e. the effective stiffness is $E_0 = 1$ and the Poisson's ratio is $\nu = 1/3$. As a result, only the yield stress limit varies according to \cref{tab:matProb}. The optimization problem is formulated as
    %
    \begin{table*}[tb]
        \centering
        \caption{Material properties of the five considered base materials (Crook et al. \cite{Crook2020}, Andersen et al. \cite{Andersen2021a}, and Fengwen and Sigmund \cite{Wang2023a})}\label{tab:matProb}
        \begin{tabular}{lcc}
            \midrule
           & $E_0 \ (\si{\giga \pascal})$ & $\sigma_0/E_0$ \\
        \midrule
        Steel & 215 & 0.002  \\
        Epoxy & 3.08 & 0.023 \\
        Pyrolytic Carbon (PC) & 62 & 0.044 \\
        Pyrolytic Carbon-Nano (PC-Nano) & 350 & 0.113 \\
        Thermoplastic polyurethane (TPU) & 0.012 & 0.333 \\
        \bottomrule
        \end{tabular}
    \end{table*}
    %
    \begin{alignat}{1}\label{eq:optProblem:LBeamYieldDomainMat:1}
        \min\limits_{\mathbf{x},\mathbf{s}}  &:   \max \left( \frac{1}{\lambda_B(\bm{\rho}^m)},\frac{1}{\lambda_Y(\bm{\rho}^m,\bar{\tilde{\bm{x}}},\bar{\tilde{\bm{s}}}^m)}, \frac{1}{\lambda_{LB}(\bm{\rho}^m,\bar{\tilde{\bm{x}}},\bar{\tilde{\bm{s}}}^m)} \right) \\ 
        & \quad + \Gamma \frac{V^e_{\Omega,struct}(\bm{s})}{V_{\Omega}}, \hspace{1.95cm} m\in\{e,i,d\}, \nonumber \\ 
        \textrm{s.t.} &:  g_{c}(\bm{\rho}^e) 
        \leq 0,  \label{eq:optProblem:LBeamYieldDomainMat:2}\\ 
         &:  g_V(\boldsymbol{\rho}^d) \leq 0, \label{eq:optProblem:LBeamYieldDomainMat:3}\\ 
        &:  \rho_e^m = \bar{\tilde{x}}_e \bar{\tilde{s}}_e^m, \hspace{2.15cm} m\in\{e,i,d\}, \ \forall e, \label{eq:optProblem:LBeamYieldDomainMat:6}\\
        &:  x_{min} \leq x_e \leq 1, \hspace{3.8cm} \forall e, \label{eq:optProblem:LBeamYieldDomainMat:7}\\
        &:  0 \leq s_e \leq 1, \hspace{4.4cm} \forall e. \label{eq:optProblem:LBeamYieldDomainMat:8}
    \end{alignat}
    %
    Here \cref{eq:optProblem:LBeamYieldDomainMat:1} is the objective which aims at maximizing the minimum of the BLFs, LBLFs and YLFs of the structure. The objective is augmented similar to the optimization problem from \cref{sec:results:yieldStressConstraint}. The compliance constraint in \cref{eq:optProblem:LBeamYieldDomainMat:2} is defined to satisfy a pre-defined value $C^*=3.45 \times 10^{-4}$ on the eroded design. The remainder of the problem is similar to the problem from \cref{sec:results:yieldStressConstraint}.
    
    \subsubsection{Optimized Designs}\label{sec:results:yieldStressConstraint:optDesigns}
\begin{figure*}[tb]
        \centering
        \graphicspath{{figures/MaterialDesigns/}}
        \makebox[\textwidth][c]{
\begingroup%
  \makeatletter%
  \providecommand\color[2][]{%
    \errmessage{(Inkscape) Color is used for the text in Inkscape, but the package 'color.sty' is not loaded}%
    \renewcommand\color[2][]{}%
  }%
  \providecommand\transparent[1]{%
    \errmessage{(Inkscape) Transparency is used (non-zero) for the text in Inkscape, but the package 'transparent.sty' is not loaded}%
    \renewcommand\transparent[1]{}%
  }%
  \providecommand\rotatebox[2]{#2}%
  \newcommand*\fsize{\dimexpr\f@size pt\relax}%
  \newcommand*\lineheight[1]{\fontsize{\fsize}{#1\fsize}\selectfont}%
  \ifx\svgwidth\undefined%
    \setlength{\unitlength}{481.88980703bp}%
    \ifx\svgscale\undefined%
      \relax%
    \else%
      \setlength{\unitlength}{\unitlength * \real{\svgscale}}%
    \fi%
  \else%
    \setlength{\unitlength}{\svgwidth}%
  \fi%
  \global\let\svgwidth\undefined%
  \global\let\svgscale\undefined%
  \makeatother%
  \begin{picture}(1,0.27658211)%
    \lineheight{1}%
    \setlength\tabcolsep{0pt}%
    \put(0,0){\includegraphics[width=\unitlength,page=1]{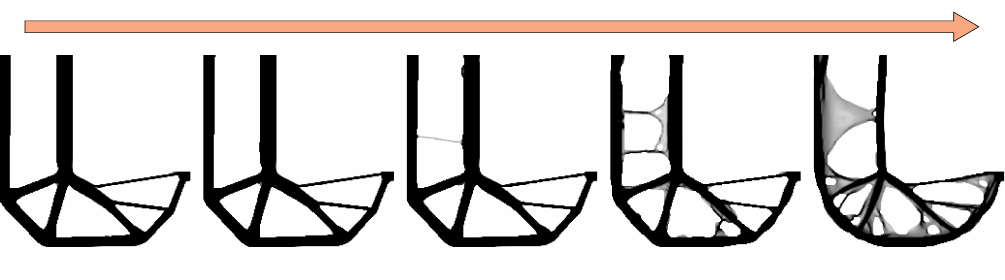}}%
    \put(0.40638843,0.26721898){\color[rgb]{0,0,0}\makebox(0,0)[lt]{\lineheight{1.25}\smash{\begin{tabular}[t]{l}Increasing $\sigma_0/E_0$\end{tabular}}}}%
    \put(0.08645837,0.0021665){\color[rgb]{0,0,0}\makebox(0,0)[lt]{\lineheight{1.25}\smash{\begin{tabular}[t]{l}(a)\end{tabular}}}}%
    \put(0.28862232,0.0021665){\color[rgb]{0,0,0}\makebox(0,0)[lt]{\lineheight{1.25}\smash{\begin{tabular}[t]{l}(b)\end{tabular}}}}%
    \put(0.49112241,0.0021665){\color[rgb]{0,0,0}\makebox(0,0)[lt]{\lineheight{1.25}\smash{\begin{tabular}[t]{l}(c)\end{tabular}}}}%
    \put(0.69387153,0.0021665){\color[rgb]{0,0,0}\makebox(0,0)[lt]{\lineheight{1.25}\smash{\begin{tabular}[t]{l}(d)\end{tabular}}}}%
    \put(0.89670785,0.0021665){\color[rgb]{0,0,0}\makebox(0,0)[lt]{\lineheight{1.25}\smash{\begin{tabular}[t]{l}(e)\end{tabular}}}}%
  \end{picture}%
\endgroup%

        \phantomsubcaption
        \label{fig:MaterialDesigns:1}
        \phantomsubcaption
        \label{fig:MaterialDesigns:2}
        \phantomsubcaption
        \label{fig:MaterialDesigns:3}
        \phantomsubcaption
        \label{fig:MaterialDesigns:4}
        \phantomsubcaption
        \label{fig:MaterialDesigns:5}
        }%
        \caption{The designs optimized specifically for the five materials: (a) Steel, (b) Epoxy, (c) PC, (d) PC-Nano, (e) TPU.}
        \label{fig:MaterialDesigns}
    \end{figure*}

    The homogenized designs of the optimizations with the five different materials are shown in \cref{fig:MaterialDesigns}. From left to right, the relative yield stress limit increases. The effect on the designs is clearly visible, with the transition from singlescale structures for the yield-dominated materials (steel and epoxy) to multiscale structures as buckling failure becomes more dominant for the high yield-limit materials (PC-Nano and TPU).
    
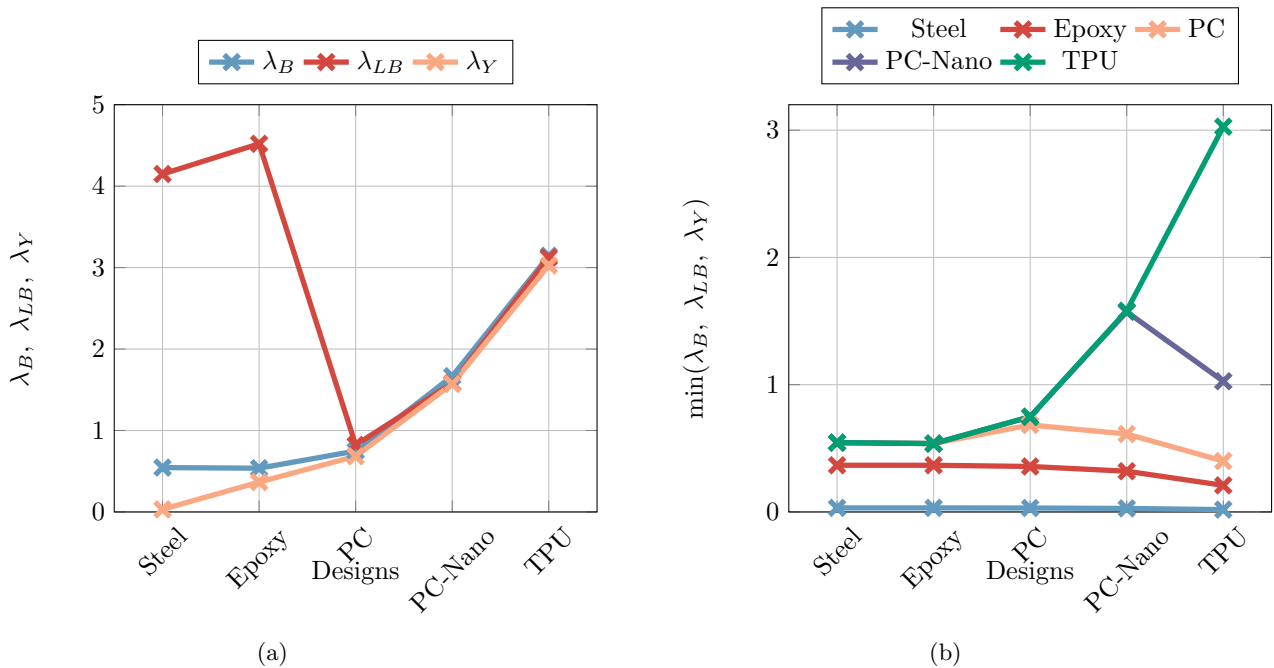
\begin{figure*}
        \centering
        \begin{subfigure}[b]{0.4\textwidth}
            \begin{tikzpicture}
    \begin{axis}[
        xlabel = {Designs},
        ylabel = {$\lambda_B, \ \lambda_{LB}, \ \lambda_Y$},
        ytick={0,1,2,3,4,5,6},
        xtick={1,2,3,4,5},
        xticklabels={Steel, Epoxy, PC, PC-Nano, TPU},
        xticklabel style={rotate=45},
        grid=major,
        xmin=0.5, xmax=5.5,
        ymin=0, ymax=5.0,
        width=8cm,   
        height=7cm,    
        legend style={
        at={(0.5,1.05)}, 
        anchor=south,    
        legend columns=3, 
    },
    ]

\pgfplotstableread{
    x  BLF LBLF YLF
    1 0.5446    4.1487    0.0319
    2 0.5372    4.5200    0.3668
    3 0.7478    0.8222    0.6837
    4 1.6664    1.5809    1.5725
    5 3.1450    3.1240    3.0269
    }\datatable

    \addplot [
        color=mycolor1,
        mark=x,
        line width = 2,
        mark size=4pt,  
    ] table [x=x, y=BLF] {\datatable};
    \addlegendentry{$\lambda_B$}

    \addplot [
        color=mycolor2,
        mark=x,
        line width = 2,
        mark size=4pt,  
    ] table [x=x, y=LBLF] {\datatable};
    \addlegendentry{$\lambda_{LB}$}

    \addplot [
        color=mycolor3,
        mark=x,
        line width = 2,
        mark size=4pt,  
    ] table [x=x, y=YLF] {\datatable};
    \addlegendentry{$\lambda_Y$}



    


\end{axis}

\end{tikzpicture}%
            \caption{}
            \label{fig:MaterialDesigns:CrossCheck:1}
    \end{subfigure}
    \hspace{1.5cm}
    \begin{subfigure}[b]{0.4\textwidth}
        \begin{tikzpicture}
    \begin{axis}[
        xlabel = {Designs},
        ylabel = {$\min(\lambda_B, \ \lambda_{LB}, \ \lambda_Y)$},
        ytick={0,1,2,3,4,5,6},
        xtick={1,2,3,4,5},
        xticklabels={Steel, Epoxy, PC, PC-Nano, TPU},
        xticklabel style={rotate=45},
        grid=major,
        xmin=0.5, xmax=5.5,
        ymin=0, ymax=3.2,
        width=8cm,   
        height=7cm,    
        legend style={
        at={(0.5,1.05)}, 
        anchor=south,    
        legend columns=3, 
    },
    ]


\pgfplotstableread{
    x Steel Epoxy PC PC-Nano TPU BLF LBLF
    1 0.0319	0.3669	0.7019	1.8027	5.3123    0.5446    4.1487
    2 0.0319	0.3668	0.7017	1.802	5.3104    0.5372    4.5200
    3 0.0311	0.3574	0.6837	1.7559	5.1745    0.7478    0.8222
    4 0.0278	0.3201	0.6123	1.5725	4.6339    1.6664    1.5809
    5 0.0182	0.2091	0.4	1.0271	3.0269    3.1450    3.1240
    }\datatable

    \pgfplotstableread{
        x Steel Epoxy PC PC-Nano TPU
        1.0000    0.0319    0.3669    0.5446    0.5446    0.5446
        2.0000    0.0319    0.3668    0.5372    0.5372    0.5372
        3.0000    0.0311    0.3574    0.6837    0.7478    0.7478
        4.0000    0.0278    0.3201    0.6123    1.5725    1.5809
        5.0000    0.0182    0.2091    0.4000    1.0271    3.0269
    }\datatabletwo

    \addplot [
        color=mycolor1,
        mark=x,
        line width = 2,
        mark size=4pt,  
    ] table [x=x, y=Steel] {\datatabletwo};
    \addlegendentry{Steel}

    \addplot [
        color=mycolor2,
        mark=x,
        line width = 2,
        mark size=4pt,  
    ] table [x=x, y=Epoxy] {\datatabletwo};
    \addlegendentry{Epoxy}

    \addplot [
        color=mycolor3,
        mark=x,
        line width = 2,
        mark size=4pt,  
    ] table [x=x, y=PC] {\datatabletwo};
    \addlegendentry{PC}

    \addplot [
        color=mycolor4,
        mark=x,
        line width = 2,
        mark size=4pt,  
    ] table [x=x, y=PC-Nano] {\datatabletwo};
    \addlegendentry{PC-Nano}

    \addplot [
        color=mycolor5,
        mark=x,
        line width = 2,
        mark size=4pt,  
    ] table [x=x, y=TPU] {\datatabletwo};
    \addlegendentry{TPU}





    


\end{axis}

\end{tikzpicture}%
        \caption{}
        \label{fig:MaterialDesigns:CrossCheck:2}
    \end{subfigure}
    \caption{Results of the deigns optimized specifically for the five materials in \cref{tab:matProb}. (a) BLF, LBLF and YLF for each design. (b) The minimum af all the load factors for each material when evaluating the designs with the other materials.}
        \label{fig:MaterialDesigns:CrossCheck}
    \end{figure*}

    The different load factors are visible in \cref{fig:MaterialDesigns:CrossCheck:1}. For the designs optimized for Steel and Epoxy, the yield limit of the material is dominant and drives the design. This effectively means that local and global buckling are neglected, thus leading to identical designs for these two materials. In the PC design, the buckling objectives start becoming active, resulting in a thin bar of intermediate density material. This increases the BLF slightly and activates the LBLF as a result of the intermediate material density. The designs optimized for PC-Nano and TPU maximize all load factors simultaneously, leading to structures with more intermediate material to increase the buckling resistance.
    %
    %
    
    To validate the tailored designs, the designs are cross-checked with the other materials. The minimum af the BLF, LBLF and YLF values are plotted in \cref{fig:MaterialDesigns:CrossCheck:2}. The figure shows that all materials provide the best performance when used on the design which is tailored for that specific material. Only Steel and Epoxy perform at the same level when used on each others design which is expected given that the designs are identical. This clearly indicates that the designs are tailored for the specific material and that the optimization problem effectively captures the material properties and utilizes them for better designs.
    
    \section{Conclusion}
    \label{sec:clonslusion}

    This study addressed the significant challenge of enhancing structural yield strength and stability in the topology optimization of multiscale structures. By integrating yield stress limits into the optimization process, the research ensures that the resulting designs meet critical material yield constraints, which is essential for practical applications.
    
    First, a two-parameter microstructure representation that offers near-optimal stiffness and stress performance was introduced. This microstructure allows for smooth de-homogenization by adjusting hole size and shape, ensuring that the designs meet the necessary structural performance criteria.
    
    Second, a methodology was developed for incorporating yield stress limits in multiscale material optimization. This approach establishes local density-dependent von Mises yield surfaces based on local yield estimates from homogenization-based analysis. Combining these local stress based YLFs with local and global buckling criteria, the method achieved topology optimized designs that consider yield and buckling failure on all levels. The effectiveness of this method was demonstrated through the L-beam example, which showed significant structural improvements over traditional singlescale methods and multiscale methods without yield stress considerations. Despite the underlying assumption of separation of scales, de-homogenization results on rather coarse length scales showed convincing agreement between homogenization results and realized de-homogenized performances. Furthermore, the test using different materials showed the importance of tailoring designs to specific material properties, leading to optimized structures that effectively utilize the material's yield strength.
    
    Overall, this research contributes to the design of safer and more reliable structures by optimizing yield strength and integrating multiscale yield stress considerations. Future work should investigate the stress evaluation inside the elements in the homogenized analyses to estimate the true maximum stresses more accurately. More accurate stress estimates can potentially provide a better correlation between homogenized and de-homogenized structures. Additionally, examining topology optimization with an upper bound to the infill density ($x_{max}<1$) using the multiscale yield constraints will force porous structures in the entire design and limit the possibility of solid material at stress concentrations, fundamentally influencing the outcome of the optimization problem.

  \section{Acknowledgments}
  This work was supported by Villum Fonden through the Villum Investigator Project “AMSTRAD” (VIL54487).
  \bibliographystyle{elsarticle-num}
  \bibliography{myBib}
\end{multicols}
\FloatBarrier
\appendix
\renewcommand{\thesection}{\Alph{section}}

\section{Two-parameter microstructure for maximum stiffness}
\label{app:twoParamterMaxStiffness}

\begin{figure*}[b!]
    \centering
    \makebox[\textwidth][c]{
            \begin{tikzpicture}
\pgfplotsset{
    scale only axis,
    xmin=0, xmax=1
}

    \pgfmathsetmacro{\p}{-1.267} 
    \pgfmathsetmacro{\pp}{4.715} 
    \pgfmathsetmacro{\ppp}{-3.927} 
    \pgfmathsetmacro{\pppp}{1.227} 
    \pgfmathsetmacro{\ppppp}{0.3806} 

    \pgfmathdeclarefunction{etaFit}{1}{%
        \pgfmathparse{\p*#1^5+ \pp * #1^4 + \ppp * #1^3 + \pppp * #1^2 + \ppppp * #1}%
        \let\pgfmathresult=\pgfmathresult
    }

    \pgfmathsetmacro{\r}{-6180} 
    \pgfmathsetmacro{\rr}{1350} 
    \pgfmathsetmacro{\rrr}{1292} 
    \pgfmathsetmacro{\rrrr}{0.4898} 
    \pgfmathsetmacro{\q}{8107 } 
    \pgfmathsetmacro{\qq}{-1491} 
    \pgfmathsetmacro{\qqq}{1467} 

    \pgfmathdeclarefunction{alphaFit}{1}{%
    \pgfmathparse{
        (\r*#1^3 + \rr*#1^2 + \rrr*#1 + \rrrr) /
        (#1^3 + \q*#1^2 + \qq*#1 + \qqq)
    }%
    \let\pgfmathresult=\pgfmathresult
    }

    \begin{axis}[
        xlabel = {$V/V_Y, \ \rho$},
        ylabel = {$\eta$},
        ymin=0, ymax=1.25,
        width=8cm,   
        height=5cm,    
        yaxis stuff style = {mycolor1},
    ]
    \addplot+[
        only marks,
        mark=*,
        color = mycolor1,
        fill=mycolor1,
        mark options={solid},
        inner sep=0.5pt,
        fill opacity=0.2,
        draw opacity=0,
     ]
    table [x index=0, y index=1, col sep=comma] {figures/YieldParamterFitMaxE/parametersMaxE.csv}; \label{pgf1:eta}

    \addplot[mycolor1,
            line width = 1,
            domain=0:1,
            samples=100]
            {etaFit(x)}; \label{pgf1:eta_hat}
\end{axis}

    \begin{axis}[
        axis y line*=right,
        axis x line=none,
        legend style={at={(0.03,0.5)},anchor=west},
        ymin=-0.65, ymax=0.22,
        ylabel={$\alpha$},
        yaxis stuff style = {mycolor2},
        ylabel style={at={(axis description cs:1.3,0.5)}, anchor=north, mycolor2},
        width=8cm,   
        height=5cm,    
      ]
      \addlegendimage{/pgfplots/refstyle=pgf1:eta}\addlegendentry{$\bar{\eta}$}
    \addplot+[
        only marks,
        mark=*,
        color = mycolor2,
        fill=mycolor2,
        mark options={solid},
        inner sep=0.5pt,
        fill opacity=0.2,
        draw opacity=0,
     ]
    table [x index=0, y index=2, col sep=comma] {figures/YieldParamterFitMaxE/parametersMaxE.csv};
    \addlegendentry{${\alpha}$}

    \addlegendimage{/pgfplots/refstyle=pgf1:eta_hat}\addlegendentry{$\hat{\bar{\eta}}(\rho)$}

    \addplot[mycolor2,
            line width = 1,
            domain=0:1,
            samples=100]
            {alphaFit(etaFit(x))};
            \addlegendentry{$\hat{\alpha}(\hat{\bar{\eta}}(\rho))$}

    \end{axis}
\end{tikzpicture}%
    }
    \caption{Parameter values and fits for a microstructure with maximum stiffness $\bar{E}$.}
    \label{fig:parameterFitE}
\end{figure*}
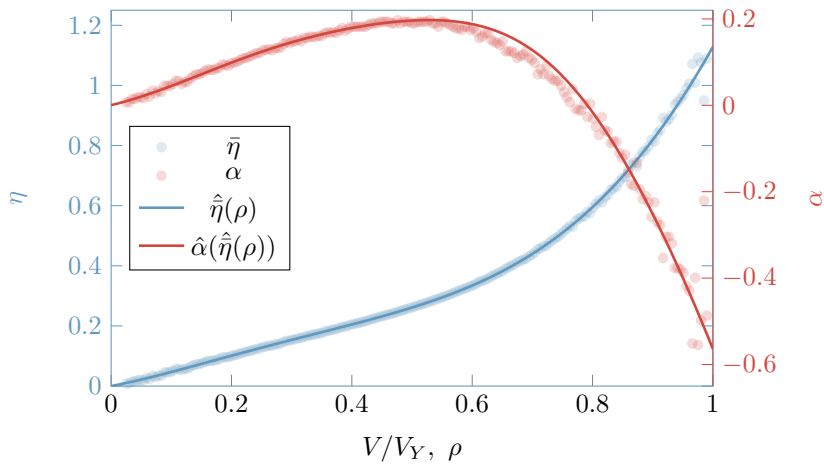

Sect. 2.2 presented the coefficients used to achieve the microstructure with the maximum yield strength. Here, the parameters needed to obtain the microstructure with the maximum stiffness are provided. The threshold fit $\hat{\bar{\eta}}$ is defined in eq. (16) with the parameters in \cref{tab:etaBarFitE}. The factor fit $\hat{{\alpha}}$ is defined in eq. (17) with the parameters in \cref{tab:alphaFitE}. \cref{fig:parameterFitE} shows that parameter values and corresponding curve fits achieving maximum effective stiffness $\bar{E}$ related to the relative density $\rho$. 
\begin{table*}[b!]
    \centering
    \caption{Values of the fitted coefficients used in $\hat{\bar{\eta}}(\rho_e)$ achieving maximum stiffness.}\label{tab:etaBarFitE}
    \begin{tabular}{rrrrrr}
     \midrule
     $\bm{p_1}$ & $\bm{p_2}$ & $\bm{p_3}$ & $\bm{p_4}$  & $\bm{p_5}$  \\
     -1.267 & 4.715 &  -3.927 & 1.227 & 0.3806  \\
     \bottomrule
     \end{tabular}
\end{table*}

\begin{table*}[b!]
    \centering
    \caption{Values of the fitted coefficients used in $\hat{\alpha}(\rho_e)$ achieving maximum stiffness.}\label{tab:alphaFitE}
    \begin{tabular}{rrrrrrr}
     \midrule
     $\bm{r_1}$ & $\bm{r_2}$ & $\bm{r_3}$ & $\bm{r_4}$ & $\bm{q_1}$ & $\bm{q_2}$ & $\bm{q_3}$ \\
     -6180& 1350 & 1292  & 0.4898 & 8107 & -1491 & 1467 \\
     \bottomrule
     \end{tabular}
\end{table*}

\end{document}